\documentclass[fleqn,usenatbib]{mnras}

\usepackage{mathptmx}

\usepackage[T1]{fontenc}
\usepackage[utf8]{inputenc}
\usepackage{ae,aecompl}

\usepackage{graphicx}	
\usepackage{amsmath}	
\usepackage{amssymb}	

\newcommand{\KR}{}
\newcommand{\newKR}{}
\newcommand{\SPZ}{}

\title[Resonant and non-resonant planetary systems in SCs]{On the survival of resonant and non-resonant planetary systems in star clusters}

\author[K. Stock et al.]{
Katja Stock$^{1}$\thanks{E-mail: katja.stock@uni-heidelberg.de}\thanks{Fellow of the International Max Planck Research School for Astronomy and Cosmic Physics at the University of Heidelberg (IMPRS-HD)},
Maxwell X. Cai$^{2}$,
Rainer Spurzem$^{1,3,4}$\thanks{Research Fellow of Frankfurt Institute for Advanced Studies.},
M.B.N. Kouwenhoven$^{5}$
\newauthor
and Simon Portegies Zwart$^{2}$
\\
$^{1}$Astronomisches Rechen-Institut, Zentrum für Astronomie der Universität Heidelberg, Mönchhofstraße 12-14, D-69120 Heidelberg,\\ Germany\\
$^{2}$Leiden Observatory, Leiden University, PO Box 9513, NL-2300 RA Leiden, the Netherlands\\
$^{3}$National Astronomical Observatories and Key Laboratory of Computational Astrophysics, Chinese Academy of Sciences,\\ 20A Datun Road, Chaoyang District, Beijing 100101, P.R. China\\
$^{4}$Kavli Institute for Astronomy and Astrophysics, Peking University, 5 Yi He Yuan Road, Haidian District, Beijing 100871,\\ P.R. China\\
$^{5}$Department of Physics, School of Science, Xi’an Jiaotong-Liverpool University, 111 Ren’ai Road, Suzhou Dushu Lake Science and\\ Education Innovation District, Suzhou Industrial Park, Suzhou 215123, P.R. China
}

\date{Accepted 2020 July 7. Received 2020 June 5; in original form 2020 February 19}

\pubyear{2020}

\begin{document}
\label{firstpage}
\pagerange{\pageref{firstpage}--\pageref{lastpage}}
\maketitle

\begin{abstract}
Despite the discovery of thousands of exoplanets in recent years, the number of known exoplanets in star clusters remains tiny.
This may be a consequence of close stellar encounters perturbing the dynamical evolution of planetary systems in these clusters. Here, we present the results from direct $N$-body simulations of multiplanetary systems embedded in star clusters containing $N=8$k, 16k, 32k, and 64k stars. The planetary systems, which consist of the four Solar system giant planets Jupiter, Saturn, Uranus, and Neptune, are initialized in different orbital configurations, to study the effect of the system architecture on the dynamical evolution of the entire planetary system, and on the escape rate of the individual planets. We find that the current orbital parameters of the Solar system giants (with initially circular orbits, as well as with  present-day eccentricities) and a slightly more compact configuration, have a high resilience against stellar perturbations. A configuration with initial mean-motion resonances of 3:2, 3:2, and 5:4 between the planets, which is inspired by the Nice model, and for which the two outermost planets are usually ejected within the first $10^5$ yr, is in many cases stabilized due to the removal of the resonances by external stellar perturbation and by the rapid ejection of at least one planet. Assigning all planets the same mass of 1\,M$_\mathrm{Jup}$ almost equalizes the survival fractions. Our simulations reproduce the broad diversity amongst observed exoplanet systems. We find not only many very wide and/or eccentric orbits, but also a significant number of (stable) retrograde orbits.

\end{abstract}




\section{Introduction}

Studies of nearby giant molecular clouds by \cite{Lada1993} suggest that stars generally do not form in isolation but also in groups or stellar associations. If clustered star formation is the rule rather than the exception, there is reason to believe that star clusters are promising targets for the detection of newborn planetary systems because star and planet formation are closely connected to each other.

Despite the large number of 4158 extrasolar planets\footnote{
As of May 2020, according to the \href{https://exoplanetarchive.ipac.caltech.edu/index.html}{NASA Exoplanet Archive}} which were detected in the last 25 yr, only around 30 planets ($<1\%$) have been detected in star clusters so far, and only one of them has been detected in a globular cluster \citep[see table 1 in][for a complete list of planet detections in star clusters \KR{and their corresponding references}]{Cai2019}. \KR{Among those planets detected in star clusters are, for example, 13 planets around 11 stars in the Praesepe (M44) cluster \citep{Quinn2012,Malavolta2016,Obermeier2016,Gaidos2017,Mann2017,Rizzuto2018,Livingston2019}, six planets in four systems in the Hyades cluster \citep{Sato2007,Quinn2014, Mann2016} with one three-planet system \citep{Mann2018}, and five single-planet systems in the M67 cluster \citep{Brucalassi2014,Brucalassi2016,Brucalassi2017}. The origin for the periodic RV variations in the giant stars IC 4651 No. 9122, NGC 2423 No. 3, and NGC  4349 No. 127, which are all located in an open cluster, is still under debate \citep{DelgadoMena2018}. \cite{Brucalassi2017} find a comparable fraction of giant planets around stars in the cluster M67 than around field stars but a significantly higher fraction of Hot Jupiters in the cluster compared to the field \citep[see also][]{Brucalassi2016}. Although the sample size in these studies is very small and statistics should therefore be interpreted with caution, the ``excess'' of Hot Jupiters found in M67 is an indication for significant dynamical perturbations from neighbouring stars on the planets in the cluster.}

Clustered environments pose a threat already for the early phases of planet formation. Protoplanetary discs may be photoevaporated by the radiation of nearby massive stars \citep[e.g.][]{Stoerzer1999, Armitage2000, Anderson2013, Facchini2016} or truncated due to close encounters \citep[e.g.][]{Clarke1993, Olczak2006, PortegiesZwart2016, Concharamirez2019}. But even when a planetary system has successfully formed without major perturbations, \KR{its dynamical fate will still be determined by the host star's position and motion inside the cluster and the properties of the cluster itself like its density (denser clusters, and especially globular clusters, tend to have a more destructive effect on planetary systems than loosely bound open clusters). Numerous studies have analysed the effect of cluster environments on planetary systems beyond the protoplanetary disc phase \citep[e.g.][]{Malmberg2007,Spurzem2009,Malmberg2011,Parker2012,Hao2013,Cai2017,Cai2018,Cai2019,Flammini2019,Fujii2019,vanElteren2019,Glaser2020}.}

\cite{Spurzem2009} presented a set of dynamical star cluster models with a large number of planetary systems (consisting of one planet) fully included into the model; they showed that there is a constant rate of planets liberated as a result of stellar encounters; they also showed that stellar encounters act like a diffusive process on planetary systems, where changes of semimajor axis and angular momentum may be directed in both ways. Depending on the details of the encounter, there is a net flux outward, giving the rate at which free-floating planets are created. There could also be a net flow to the inner boundary, i.e. planets accreted onto the central star, which was not discussed in their paper. \cite{Li2015} followed another approach – a Monte Carlo model, in which many thousands of encounters of single objects (single and binary stars) with planetary systems were modelled. They were able to cover a parameter space substantially larger than that of \cite{Spurzem2009}. However, Monte Carlo models suffer from the inaccuracy in the stochastic selection of encounter parameters (the impact parameter and the velocity at infinity). In figs~1 and~2 in \cite{Spurzem2009} one can see that the real distribution of these parameters in a star cluster differs from a random selection, covering the available phase space equally.

\begin{table}
\centering
\caption{Initial conditions for the star cluster simulations.}
\label{tab:SC_IC}
\begin{tabular}{lcccc}
\hline
Star cluster & 8k & 16k & 32k & 64k\\
\hline
Number of stars & 8000 & 16\,000 & 32\,000 & 64\,000 \\
Total mass (M$_\odot$) & 4073 & 7939 & 16\,302 & 32\,619 \\
Half-mass radius (pc) & 0.78 & 0.78 & 0.78 & 0.78 \\
Central density (M$_\odot\mathrm{pc}^{-3}$) & 3906 & 6813 & 13\,852 & 25153 \\
Initial tidal radius (pc) & 22.58 & 28.20 & 35.84 & 45.16\\
\hline
\end{tabular}
\end{table}

\begin{figure}
    \centering
    \includegraphics[width=0.47\textwidth, trim= 0.25cm 0.15cm 1.6cm 1.25cm,clip]{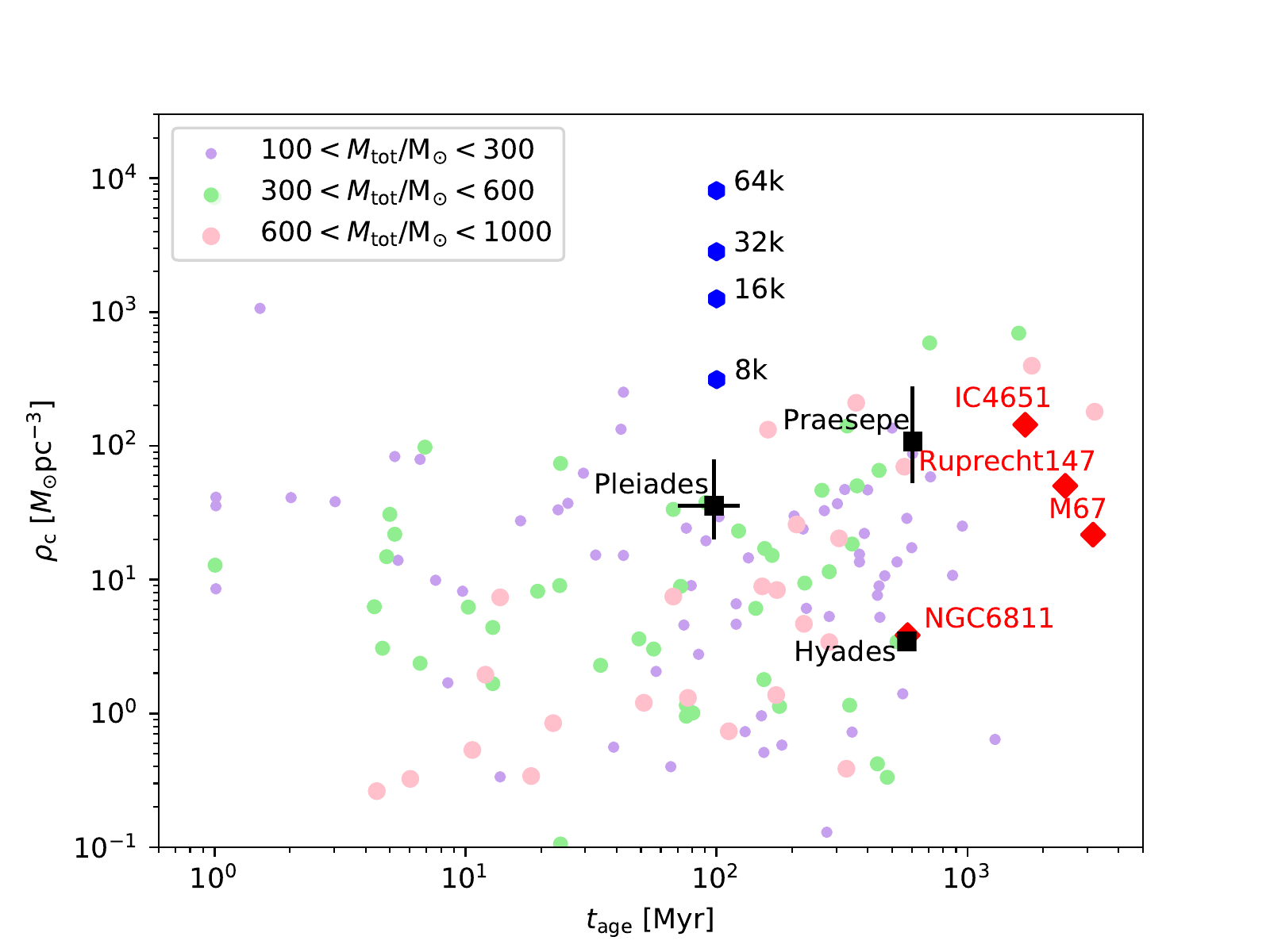}
    \caption{\KR{Central density ($\rho_\mathrm{c}$) as a function of the cluster age $\tau_\mathrm{age}$, for our four simulated clusters (blue hexagons) at a simulation time of 100\,Myr and the observed clusters from fig.~1 in~\protect\cite{Fujii2019}.}} 
    \label{fig:comparison_Fujii_Hori_2019}
\end{figure}

\SPZ{In the work of \cite{vanElteren2019}, they adopted a different approach, in which planetary systems were integrated together with the stars in the cluster. To reduce the computational burden, planets in one system were not affecting the orbits of planets in another system. This still led to an enormous computational burden, which resulted in a rather limited parameter study.}

Note that also very low-mass particles, such as planetesimals (\SPZ{asteroids} and Kuiper belt objects) or comets (Oort cloud objects) are subject to these encounters \citep[see e.g.][]{Veras2020}. \SPZ{A characterization of the importance of such close encounters on planetary systems with debris discs is presented in \cite{PortegiesZwart2015}.}
This process can lead to flybys of interstellar objects \citep{Torres2019} or to the capture of cometary objects into young planetary systems \citep[e.g.,][]{Kouwenhoven2010, Perets2012, Wangzheng2015, Shu2020}. This work is also closely related to a – somewhat less comprehensive – study by \cite{Hands2019}. 
\SPZ{It was suggested that the extraordinary asteroids 90377~Sedna was abducted from the debris disc of another star in such a close encounter \citep{Jilkova2016}.}
The identification of `Oumuamua and 2I/Borisov as possible interstellar objects in our Solar system has received much attention recently, and is connected to the idea that young planetary systems are sources of free-floating comets or planetesimals \citep[see, e.g.,][and references therein]{Zhengwang2015,PortegiesZwart2018,Hands2019,Bannister2019,Pfalzner2019}.

In this work, we study the effect of close stellar encounters on the dynamical architectures of planets that are born around stars in star clusters. We pay particular attention to the dependence on the initial orbital configuration of a planetary system before the first encounters with neighbouring stars take place. \KR{Our work differs from earlier works in several aspects. (i) We do not only focus on the effect of one single encounter on planetary systems but instead investigate the cumulative effect of several encounters on planetary systems by following their dynamical evolution during a significant fraction of time which they spend in the cluster. This again allows us to compare the distribution of orbital parameters at the end of our simulations with actual observed properties of planetary systems that are in conflict with current planet formation theories (e.g. eccentric or retrograde orbits). (ii) Our $N$-body approach enables a realistic representation of encounters between cluster members, while many previous works use a Monte Carlo approach which typically suffer from inaccuracy by randomly selecting encounter parameters equally from the available parameter space. (iii) Using a hybrid $N$-body code allows us to put every planetary system in different initial configurations while the host star's trajectory through the cluster and thus also all external perturbations on the planetary system are the same for the different system architectures.}

This paper is organized as follows. Section~\ref{sec:methods} describes the computational approach of the simulation of planetary systems embedded in star clusters and specifies the initial conditions for the star cluster simulation and the simulation of the planetary systems. In Sec.~\ref{sec:results} we present the results of our simulations which are then discussed and summarized in Sec.~\ref{sec:conclusion}.

\section{Methods and initial conditions}\label{sec:methods}

\subsection{Computational approach}\label{sec:computational_approach}

Planetary systems evolve through secular evolution, the orbits of planets being relatively stable for millions, and sometimes tens of billions of orbits. Secular evolution is provided by mutual gravitational interaction between the planets, as well as by external perturbation through passing stars, in a star cluster or in the Galaxy. However,  stellar clusters evolve differently, namely through two-body relaxation and few-body encounters. Orbits of stars in the system are changed by these processes in less than a single orbital time-scale. The dynamical evolution of star clusters also exhibits deterministic chaos, so that slightly different initial conditions can lead to exponentially diverging outcomes in phase space within less than one orbital time \SPZ{\citep[see e.g.][]{Miller1964, Quinlan1992,Boekholt2020}}.

Therefore, a combined simulation of planetary systems in star clusters is a challenge. The challenge lies not so much in the different time-scales or hierarchical nature of some objects (in this sense, close stellar binaries and planetary system are quite similar); rather the problem is to accurately follow resonant and secular effects in the internal evolution of planetary systems. This is why we simulate star cluster and planetary systems using different simulation codes. This is feasible because we assume that the neighbouring stars in the cluster affect the planets, but the planets have a negligible influence on the stellar kinematics.

Although currently the decoupled, combined simulations of planetary systems and star clusters as described earlier are state of the art, and fully coupled dynamical simulations of planetary systems in star clusters have only been carried out for single planetary systems \citep[e.g.,][]{Spurzem2009}, in the future more development on that side would be important. Using the current \texttt{LPS} algorithm (see below) neglects the potential effect of more distant perturbers and also tidal forces of the entire star cluster on the planetary system. Also very massive bodies being further away (e.g. stellar or intermediate mass black holes) could have an impact on planetary systems which are not taken into account here. 

We first simulate the stellar population in the star cluster using \texttt{NBODY6++GPU} \citep{Wang2015,Wang2016} and integrate the motion of its members inside the cluster using the Hermite scheme. \texttt{NBODY6++GPU} is a follow-up version of \texttt{NBODY6} \citep{Aarseth1999} and \texttt{NBODY6++}, and has a significant speedup due to the usage of graphical processing units (GPUs) and parallelization of tasks through a message passing interface (MPI). All required information such as mass, position, velocities, acceleration, and the first time derivative of the acceleration of all cluster members in our simulation are stored at a high time resolution using the ``block time-step'' (BTS) storage scheme \SPZ{\citep{Faber2010,Farr2012,Cai2015}}. This scheme allows the reconstruction of stellar encounters in details when planetary systems are assigned to single stars in the cluster at a subsequent step (see Sec.~\ref{sec:ic_planetary_systems}). The data are stored in \texttt{HDF5}\footnote{\url{https://www.hdfgroup.org/}} format to enable high-performance parallel access to the data.

The dynamical evolution of the planetary systems is simulated using the \texttt{LonelyPlanets Scheme} (\texttt{LPS}). It is based on the \texttt{AMUSE} framework \citep{PortegiesZwart2011,AMUSEbook} and uses \texttt{rebound} \citep{Rein2012} to integrate the planets. Before integrating the planets using the \texttt{IAS15} integrator \citep{Rein2015}, all encounters with the next five neighbouring stars are derived by interpolating the data of the corresponding stars from the BTS data \citep[see][for further explanations]{Cai2017,Cai2019}.

\subsection{Star cluster simulations}\label{sec:ic_star_cluster}

The simulated star clusters in this work contain 8000, 16\,000, 32\,000, and 64\,000 stars. We adopt the \cite{Kroupa2001} initial mass function in the mass range of 0.08-100\,M$_\odot$. The stars have an expected average mass of 0.509\,M$_\odot$. We draw the initial positions and velocities for the stars in our clusters from the \cite{Plummer1911} model. The initial half-mass radius for all clusters is $r_\mathrm{hm}=0.78\,\mathrm{pc}$. We do not include primordial mass segregation and we do not include primordial binary systems. All initial parameters for the star cluster simulations are listed in Tab.~\ref{tab:SC_IC}.
\KR{It should be noted that these values are initial cluster properties. After a short phase of core collapse the clusters rapidly expand and the central densities decrease significantly. Simulating the clusters for $100-250$\,Myr leads to central densities that correspond to those of observed star clusters. Figure~\ref{fig:comparison_Fujii_Hori_2019} shows the central density of our simulated clusters after a simulation time of 100\,Myr in comparison to the actual observed clusters from fig.~1 in \cite{Fujii2019}.
The central density is not comparable to the typical density our planetary systems experience during their life in the cluster, and due to the onset of mass segregation it is unlikely that our 1\,M$_\odot$ host stars remain in the small but dense core of a Plummer model cluster for a long time. The vast majority of all systems experience moderate stellar densities of up to a few hundred M$_\odot$\,pc$^{-3}$.}

\KR{The encounter time-scale \citep[see equation~3 in][]{Malmberg2007} determined for our clusters is comparable to $\tau_\mathrm{enc} \approx 2.4\,$Myr given in \cite{Malmberg2007}. Instead of using $m_\mathrm{t}=1\,\mathrm{M}_\odot$ for the total mass of the stars involved in the encounter as \cite{Malmberg2007} do, we use a value of $m_\mathrm{t}=1.5\,\mathrm{M}_\odot$ based on the assumption that our 1\,M$_\odot$ host stars encounter stars with the average mass of $\sim 0.5$\,M$_\odot$. We set $r_\mathrm{min}= 1000$\,au as the encounter distance to ensure comparability with \cite{Malmberg2007}. For the smallest cluster, we obtain a value of $\tau_\mathrm{enc} \approx 3.2\,$Myr, and $\tau_\mathrm{enc} \approx 1.1\,$Myr for the largest cluster. This corresponds to encounter rates of 0.3 (smallest cluster) and 0.9 (largest cluster) encounters per star per Myr.}

\begin{figure}
\includegraphics[width=0.47\textwidth, trim= 1.cm 0.9cm 2.5cm 1.5cm,clip]{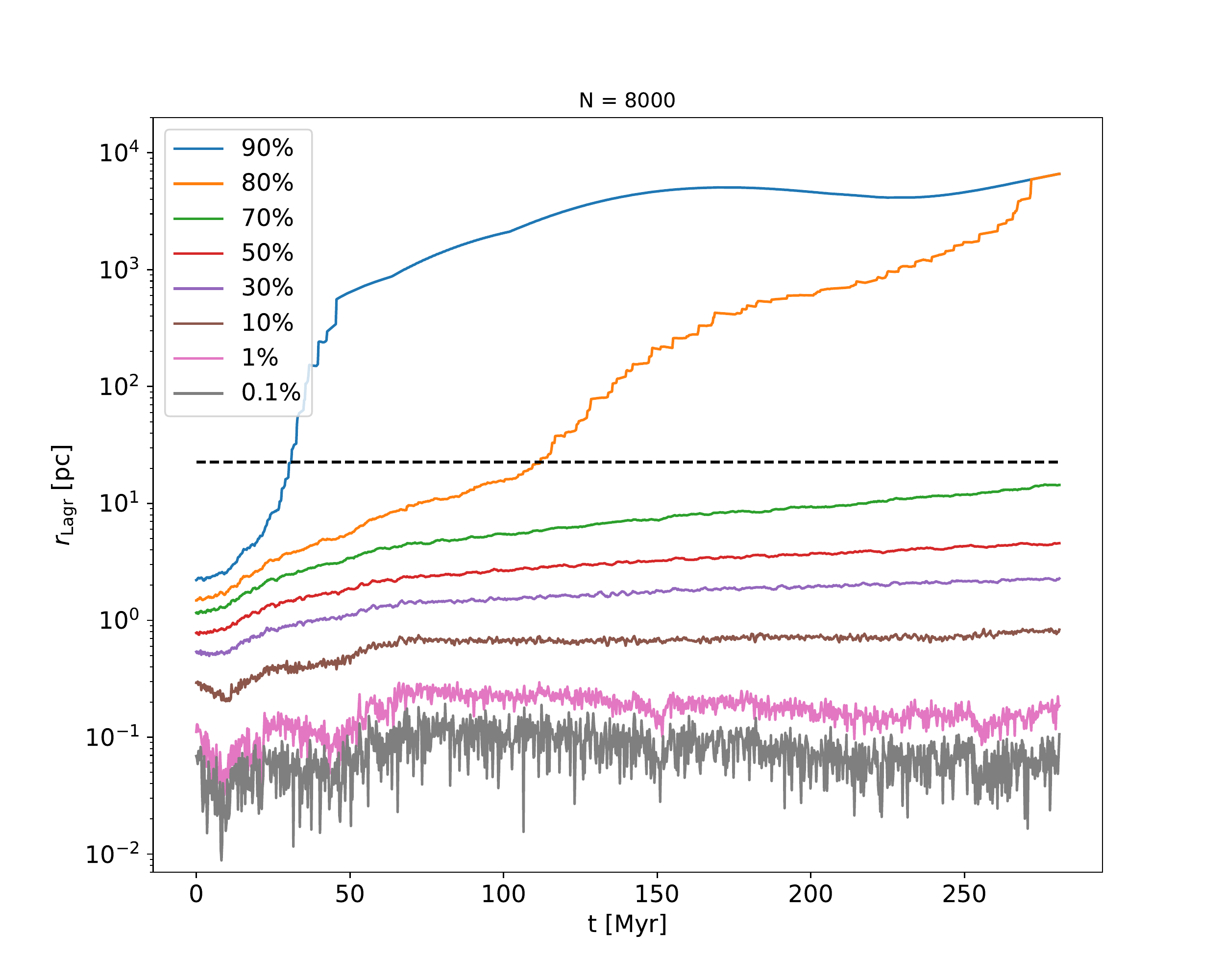}
\caption{Lagrangian radii $r_\mathrm{Lagr}$ of the 8k star cluster, containing the indicated fraction of mass, as a function of time. The black dashed curve shows the initial tidal radius $r_\mathrm{tide}$.}
\label{fig:Rlagr8k}
\end{figure}

The Lagrangian radii containing different fractions of the total cluster mass as a function of time are shown in Fig.~\ref{fig:Rlagr8k} for the 8k cluster and give an overview of the evolution of the entire star cluster. For comparison, the initial tidal radius $r_\mathrm{tid}$ is plotted as well. The half-life of the cluster is defined as the time at which the 50\% Lagrangian radius (half-mass radius) and the tidal radius are equal.

We use the standard definition\footnote{Note that this definition of the tidal radius is an operational one, used for example in our $N$-body code; other definitions use the truncation of the density profile \citep{King1962} or the distance between the Lagrange point and the cluster centre \citep[see e.g.][]{Just2009}. These definitions differ from ours by a numerical factor of order unity.} of the tidal radius as
\begin{equation}
r_\mathrm{tid} = R_\mathrm{G} \bigg(\frac{M_\mathrm{cl}}{M_\mathrm{G}}\bigg)^{\frac{1}{3}} \quad ,
\end{equation}
where $R_\mathrm{G}$ and $M_\mathrm{G}$ are the distance to the Galactic centre and the mass of the Galaxy contained inside $R_\mathrm{G}$; $M_\mathrm{cl}$ is the star cluster mass. Our star clusters gradually lose mass over time due to stellar evolution (see below) which results in a shrinking tidal radius over time.

Near the tidal radius stars are typically only marginally bound to the cluster, and may escape from the cluster into the tidal tails. In reality the situation is much more complex, since stars escape through Lagrangian points, and not all stars with positive energy (or outside $r_\mathrm{tid}$) escape immediately, some of them may be retained by the cluster. This process is neatly described in the study of \cite{Ernst2008}.

Our star clusters are assumed to orbit the Galactic centre in the solar neighbourhood, wherefore the tidal forces of the galaxy on the cluster are the same as for the solar neighbourhood \citep{Heisler1986}. The formation of tidal tails is observed in our simulations. We do not remove stars from our simulations even when their position is $r \gg r_\mathrm{tid}$. Therefore, we still keep track of the motion of stars in our simulation that have physically already left the cluster. Hence, the cluster dissolves faster than Fig.~\ref{fig:Rlagr8k} suggest. We assume that the clusters will have reduced their central density significantly after roughly 100\,Myr. For example, in the 8k cluster, almost 20\% of the stars are already beyond the tidal radius after 100\,Myr and can be considered to have left the cluster. Therefore, we simulate the cluster environment of our planetary systems only for this time span as most strong encounters will have occurred by that time.

The star cluster simulations with \texttt{NBODY6++GPU} include stellar evolution of single and binary stars --– they follow the evolution of masses and radii of all objects according to the recipes described in \citet[and earlier citations of Hurley therein]{Hurley2005}. Since we start without primordial binary stars, binary systems are rare --- only a few dynamically formed binaries are found. The stellar evolution is implemented in the form of parametrized lookup tables; any mass-loss of stars or from binaries is assumed to leave the cluster instantaneously; mass transfer in a binary is approximately followed. The reader interested in more details could have a look into the DRAGON (million body) simulations \citep{Wang2016}. In recent years the stellar evolution prescriptions for $N$-body simulations are undergoing considerable changes, see for example \SPZ{\cite{Khalaj2015}, \cite{Spera2015}, and \cite{Banerjee2019}} for an overview. The updates according to that paper are now also available in \texttt{NBODY6++GPU}, but have not yet been used for the simulations of this paper. Note that we select in the \texttt{LPS} scheme only host stars for planets which are close to one solar mass – therefore these systems are not subject to any changes due to stellar evolution, given the relatively short time of simulation used here. In future models, we could also initialize planets around more massive stars, which would undergo changes due to stellar evolution (mass-loss due to expansion of the host star on the AGB leads to a loss of planets or wider orbits of those remaining). 

\begin{table*}
\centering
\caption{Initial orbital parameters of our planetary systems in the different configurations from~\protect\cite{Li2015}.}
\label{tab:planets_IC}
\begin{tabular}{llllllll}
\hline
Configuration & \multicolumn{3}{c}{Common parameters} &  Jupiter & Saturn & Uranus & Neptune\\
\hline
\hline
Standard & $e=0$ & $i=0^\circ$ & & $a=5.20$\,au & $a=9.54$\,au & $a=19.19$\,au & $a=30.08$\,au\\
\hline
Compact & $e=0$ & $i=0^\circ$ & & $a=5.20$\,au & $a=8.67$\,au & $a=14.4$\,au & $a=24.1$\,au\\
\hline
Resonant & $e=0$ & $i=0^\circ$ & & $a=5.88$\,au & $a=7.89$\,au & $a=10.38$\,au & $a=12.01$\,au\\
\hline
Eccentric \#1	& & $i=0^\circ$ & & $a=5.20$\,au & $a=9.54$\,au & $a=19.19$\,au & $a=30.08$\,au\\
 & & & & $e=0.049$ & $e=0.057$ & $e=0.045$ & $e=0.011$ \\
\hline
Eccentric \#2 & $e=0.1$ & $i=0^\circ$ & & $a=5.20$\,au & $a=9.54$\,au & $a=19.19$\,au & $a=30.08$\,au\\
\hline
Massive	& $e=0$ & $i=0^\circ$ & $m_\mathrm{pl}=1$\,M$_\mathrm{Jup}$ & $a=5.20$\,au & $a=9.54$\,au & $a=19.19$\,au & $a=30.08$\,au\\
	
\hline
\end{tabular}
\end{table*}

\subsection{Planetary system simulation}\label{sec:ic_planetary_systems}

We aim to investigate how the initial configuration of the planetary systems affects the dynamical evolution of the planets that are born around stars in clustered environments and how it affects the likelihood of the individual planets to survive the first tens of millions of years in such a destructive environment. For this purpose, we adopt the six different initial configurations of \cite{Li2015} as starting positions for the planets in our simulations (see Tab.~\ref{tab:planets_IC}).

\cite{Li2015} study scattering encounters between Solar system analogues and passing stars (single stars and binary systems) and determine cross-sections for the disruption of these planetary systems. Their planetary systems contain the four Solar system giants Jupiter, Saturn, Uranus, and Neptune with their present-day masses. In the ``standard configuration'' of \cite{Li2015}, they use the current semimajor axes of the planets but they assign circular and co-planar orbits. Inspired by the Nice model \citep{Gomes2005}, \cite{Li2015} use two more compact configurations in which the three outer planets are closer to Jupiter. The first one is referred to as ``compact configuration''. \KR{Although these planetary systems are tightly packed, this configuration is fully stable over 100\,Myr.} In the second one, the four planets are in mutual mean-motion resonance (MMR), wherefore this configuration is called ``resonant configuration''. In this configuration, Jupiter/Saturn and Saturn/Uranus are each in a 3:2 MMR while Uranus and Neptune are in a 5:4 MMR. See \cite{Li2015} and the references therein for a further discussion of this initial state. The initial orbital angles in this configuration play a key role in the question whether not only this system is stable for a certain period of time but also the resonance angles librate for similar period of time. In our simulations, we can fulfil the stability and resonance criterion usually long enough until the first encounters of neighbouring stars start to disturb the planetary systems and break the resonances between the planets. However, it should be mentioned that this resonant configuration is generally highly unstable due to its compactness, and usually at least one of the outer planets is ejected rapidly when the initial orbital parameter are not chosen properly.

Furthermore, \cite{Li2015} use two eccentric configurations (referred to as ``Eccentric \#1'' and ``Eccentric \#2''). In the first eccentric case, the planets start again at their current semimajor axes but with their actual eccentricities (instead of circular orbits as in the standard configuration). In the second eccentric configuration, all four planets have initial eccentricities of $e=0.1$. \KR{While the first eccentric configuration is fully stable over 100\,Myr, \newKR{Neptune is ejected in the second eccentric configuration after 5\,Myr if we place the system in isolation. Therefore, the second eccentric as well as the resonant configuration both contain an internal instability leading to a higher vulnerability against external perturbations.}} The sixth investigated configuration in \cite{Li2015} is referred to as ``massive configuration'' in which all planets have Jovian masses instead of their actual masses. \KR{Despite the large masses of all four planets, the configuration is stable for at least 100\,Myr.}

For all these six configurations, we distribute 200 identical planetary systems around those stars in the cluster whose masses are closest to 1\,M$_\odot$. The host stars within one cluster simulation are therefore the same for each configuration. This allows us to work out the differences in vulnerability in the clustered environment between those initial configurations due to the different positions of the host stars in the cluster. The number of 200 planetary systems per cluster and per configuration is a compromise between computational costs and the possibility to do proper statistics about our sample. Since we simulate 200 planetary systems in six different configurations in all four star clusters, we have a total number of 4800 different planetary system simulations. On grounds of efficiency, our simulations are therefore carried out using the simulation monitor \texttt{SiMon} \citep{Qian2017}.

Each planetary system is integrated for 100\,Myr (as discussed in Sec.~\ref{sec:ic_star_cluster}). Planets that are excited to an eccentricity $e > 0.99$ are considered as having been ejected from the system and are removed from the simulation. The mass-loss of the $\sim$1\,M$_\odot$ host stars is negligible during the main-sequence phase and especially during the first 100\,Myr which is why it is not taken into account for the dynamical evolution of the planetary systems.

\section{Results}\label{sec:results}

\subsection{Fractions of surviving planets}\label{sec:survival_rates}

\begin{figure*}
	\includegraphics[width=2\columnwidth]{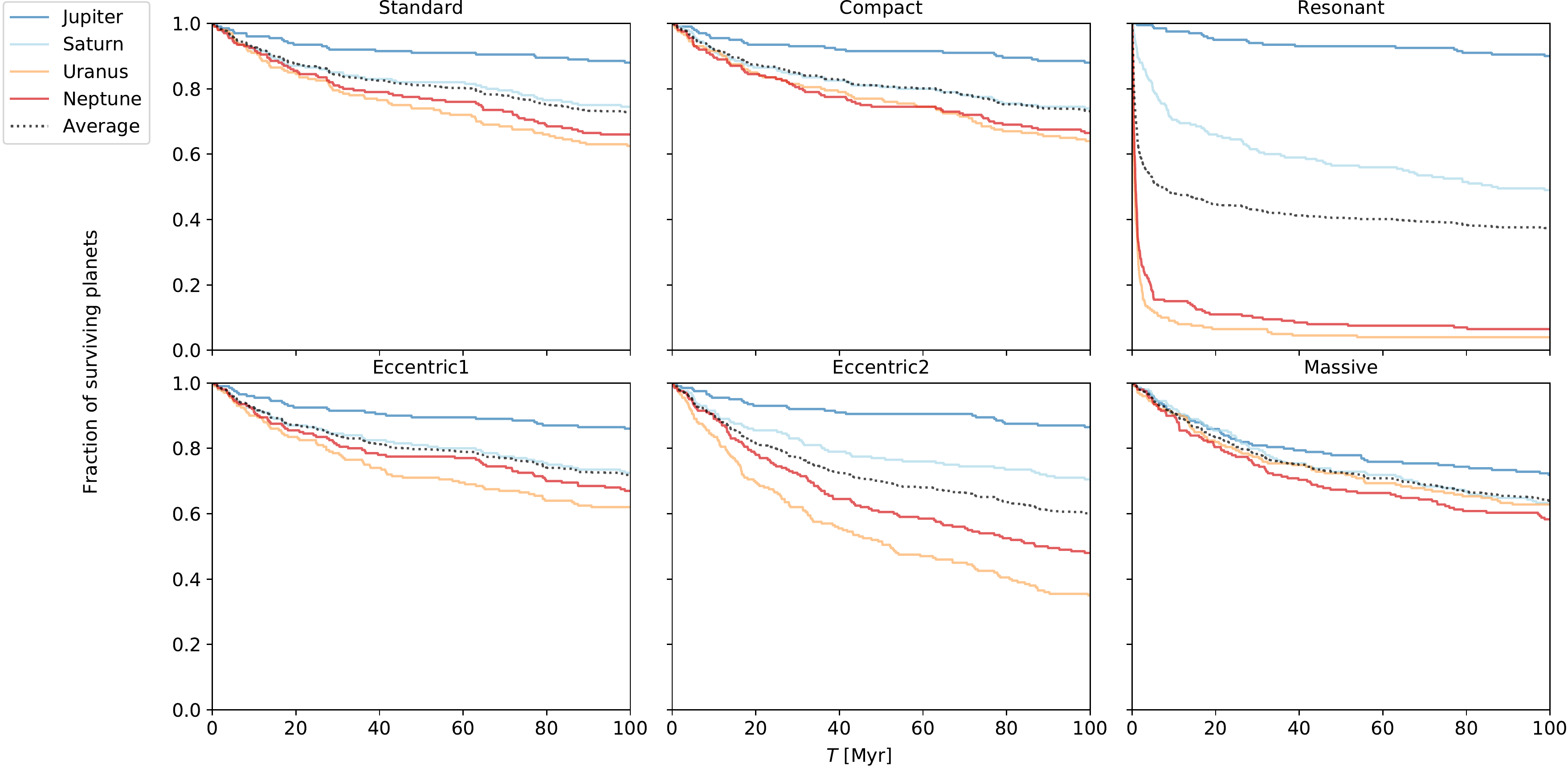}
    \caption{The survival fractions for the Solar system giant planets as a function of time for the six different initial configurations in a 16k Plummer model star cluster. The black dotted curves represent the overall survival fraction averaged over the four planets.}
    \label{fig:survival_rate_16k}
\end{figure*}

For each configuration, we simulate 200 identical planetary systems and distribute them around $\sim$1\,M$_\odot$ host stars. As we do not include primordial mass segregation in our clusters the positions of the host stars \KR{(and therefore the stellar densities the planetary systems experience)} in our clusters are random. However, \KR{to ensure comparability between} the different initial configurations we use the same 200 host stars for all planetary systems of the same cluster.

In this work we define a planet having ``survived'' when it has not been ejected from the planetary system during the course of the simulation. This means that the planet's eccentricity  has been $e\leq 0.99$ for at least 100\,Myr. For the determination of the survival fraction of a certain kind of planet we average over all 200 planets of the same type in the same star cluster.

\begin{figure*}
	\includegraphics[width=2\columnwidth]{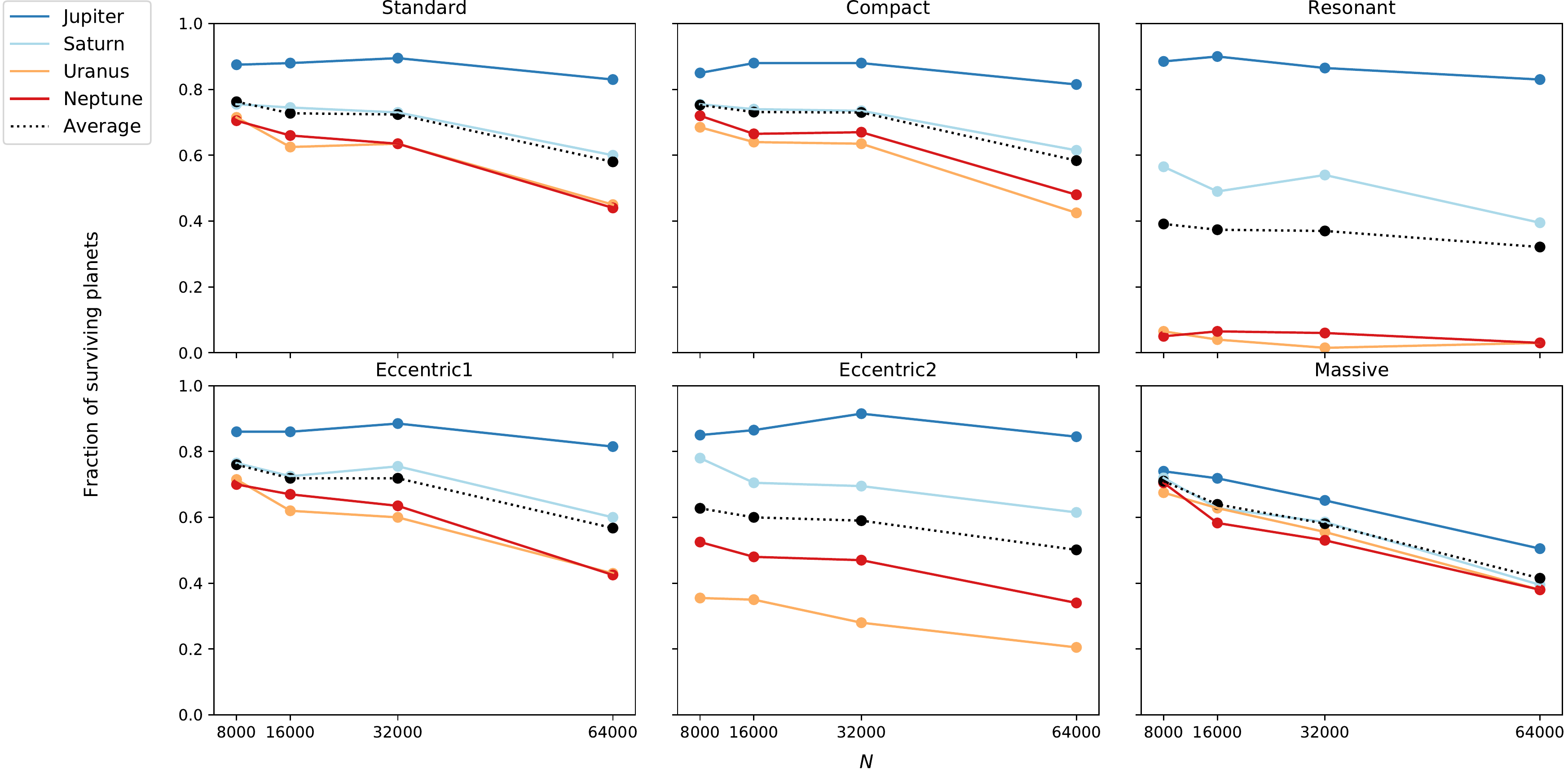}
	\caption{The survival fractions for the Solar system giant planets as a function of the number of stars, $N$, in the host star cluster at $t=100$\,Myr.}
    \label{fig:survival_rate_N}
\end{figure*}

An inspection of the survival fraction as a function of time for the four different planets in our systems reveals large differences between the initial configurations. Figure~\ref{fig:survival_rate_16k} shows the survival fraction for all six configurations as a function of time for the 16k cluster. In Fig.~\ref{fig:survival_rate_N}, the fraction of surviving planets after 100\,Myr is plotted against the number of stars, $N$, in the host star cluster, for the six initial orbital configurations.

In all of our simulations, Jupiter is the planet with the highest survival probability. The reason for this is two-fold: Jupiter is not only the most massive planet (in five of our six configurations) but it is also the innermost planet, so that its binding energy is by far the largest. The same reasoning explains why Saturn is usually the second most resistant planet. Although Neptune is the outermost planet, its binding energy is somewhat larger than that of Uranus due to its larger mass. This is why in most of our simulations Neptune is slightly more likely to survive than Uranus.

The survival fractions after 100\,Myr for the planets in standard configuration in the 16k cluster are 88.0\%, 74.5\%, 62.5\%, and 66.0\% for Jupiter, Saturn, Uranus, and Neptune, respectively. Starting the planets in the 16k cluster in a more compact configuration does not change these values significantly as can be seen in Fig.~\ref{fig:survival_rate_16k}. Also the first eccentric configuration in which the planets were assigned their true eccentricities does not differ significantly from the standard and compact case. The overall survival fraction (averaged over all four planets) is around 72\% for these three configurations in the 16k cluster.

While the differences in the survival fractions of the standard and compact configurations are negligible, the survival fractions in the resonant case differ significantly from those in the compact configuration. Only the fraction of surviving Jupiters is comparable to the other configurations and is 90.0\% in the 16k cluster. For Saturn, the survival fraction decreases from 74.5\% in the standard case to 49.0\% in the resonant configuration. However, the percentage of surviving Uranus- and Neptune-like planets is much lower and is only 4.0\% and 6.5\%, respectively. The overall survival fraction in the resonant case is only 37.4\% which is the lowest value for all six configurations in the 16k cluster. Although all planets have initially circular orbits, the effect of planet--planet interaction in this configuration is very destructive. Due to the compactness of the planetary system, the system is only long-term stable on time-scales of several ten thousand years. The first encounters have usually already occurred at that time, removing the system from resonances and exciting the orbital parameters of some or all planets. Uranus and Neptune are the most vulnerable planets in this configuration. In none of our simulations, all four planets survived. Usually either Uranus or Neptune (or both) is ejected latest after several million years. Only in 2 out of 800 simulations of the resonant case, Uranus and Neptune survived together. In both cases, Saturn is ejected within the first two million years.

In the second eccentric configuration, the survival fractions of Jupiter and Saturn are relatively unaffected by the larger initial eccentricities of all four planets. Only for Uranus and Neptune, the differences are significant compared to the standard, compact, and first eccentric case. In the 16k cluster, the survival fractions after 100\,Myr drop down to 35\% and 48\% for Uranus and Neptune, respectively. Although Neptune is the outermost planet, it has a significant higher chance to survive in this eccentric planetary system than Uranus. In this configuration, Uranus' fate is mainly determined by secular evolution. Since the planets already have an initial eccentricity of $e=0.1$, it requires less angular momentum transfer to another star to trigger destructive interactions between the planets. Due to its position between Saturn and Neptune and the fact that it has the smallest mass, Uranus is easily excited to highly eccentric orbits which often leads to the ejection of the planet.

The fractions of surviving planets in the massive configuration reveal not only the importance of the planetary mass during stellar encounters but also its role during secular evolution. The overall survival fraction in the 16k cluster drops from 72.8\% in the standard configuration to 63.9\% in the massive configuration. While Jupiter had by far the largest likelihood for survival in the other configurations, the differences between the planets in the massive configuration are significantly smaller. The survival fraction for Jupiter in the 16k cluster is 88.0\% in the standard and compact case, but only 71.9\% in the massive configuration. For Saturn, Uranus and Neptune the survival rates in the massive configuration are all around 60\% in the 16k cluster.

Our simulated clusters all have the same initial half-mass radius but differ in central density. Therefore, the survival fractions for the different configurations also depend on the number of stars in the host cluster. In general, the survival fractions for the different planets decrease with increasing stellar density due to an increasing number of close encounters between cluster members. However, the effect of an increasing stellar density is larger on the outer planets of the system since they are more easily liberated by another star due to their smaller gravitational binding energy. While the survival fractions for Jupiter, Saturn, Uranus, and Neptune in the standard configuration in the 8k cluster are 87.5\%, 75.5\%, 71.5\%, and 70.5\%, respectively, these values decrease to 83.0\%, 60.0\%, 45.0\%, and 44.0\%, respectively, in the 64k cluster (see Fig.~\ref{fig:survival_rate_N}).

\subsection{Comparison of the survival fractions with previous studies}\label{sec:comparison}

In table 2 in \cite{Li2015}, the authors provide their ejection cross-sections in units of $\mathrm{au}^2$. In order to normalize this to obtain an escape or survival fraction, one needs to know the maximum impact parameter chosen in their models. However, their maximum impact parameter is variable, depending on parameters (e.g. 10 times the semimajor axis of a stellar binary which encounters a planetary system, but not more than 1000 au). To be able to compare our results with those of \cite{Li2015}, we adopt $p_\mathrm{max}=1000$\,au for the normalization of their cross-sections.
Table~\ref{tab:survival_rates_comparison} lists our survival fractions in percentage after integrating the planetary systems in the 8k cluster in four of the six different initial configurations for 100\,Myr. We assume that our smallest cluster is most similar to the cluster environment simulated in \cite{Li2015}. However, our models are different in three aspects --- (1) the distribution of impact parameters and relative velocities of encounters is very different to the one assumed in Monte Carlo simulations of encounters as they did; (2) \cite{Li2015} stop the planetary system model after the encounter, while we continue all planetary systems for the entire simulation time of 100\,Myr and find many delayed unstable systems, which reduce the survival fraction; (3) our simulations take into account the cumulative effect of several encounters. Due to these differences, we find much more ejections of planets in our simulations and have significantly smaller survival fractions for each planet type than \cite{Li2015}.

\begin{table*}
    \centering
    \caption{Survival fractions (in percent) at $t=100$\,Myr in the 8k cluster, in comparison to the results of~\protect\cite{Li2015}.}
    \label{tab:survival_rates_comparison}
    \begin{tabular}{lccccccccc}
    \hline
    & Jupiter & Saturn & Uranus & Neptune & Jupiter & Saturn & Uranus & Neptune \\
    \hline
    \hline
    & \multicolumn{4}{c}{8k} & \multicolumn{4}{c}{Li \& Adams (2015)}\\
    \hline
    Standard & 87.5 & 75.5 & 71.5 & 70.5 & 98.5 & 96.6 & 92.8 & 88.7\\
    Compact & 85.0 & 75.5 & 68.5 & 72.0 & 98.2 & 96.7 & 94.3 & 90.6\\
    Resonant & 88.5 & 56.5 & 6.50 & 5.0 & 97.6 & 96.0 & 93.9 & 94.0\\
    Massive & 74.0 & 72.0 & 67.5 & 70.5 & 97.6 & 96.2 & 92.2 & 89.5\\
    \hline
    \end{tabular}
\end{table*}

The ejection of one or several planets can occur either during or directly after the encounter (prompt ejection), or at a later time due to secular evolution (delayed ejection). For the standard case in the 16k cluster, we find a fraction of $\sim 60\%$ prompt ejections and $\sim 40\%$ delayed ejections (see Table~\ref{tab:prompt_vs_delayed_ejection} for a distinction between the different planet types). However, both events (but especially the latter case) are not well defined in our simulations since the planetary systems are continuously perturbed by other stars. In many cases, where planetary systems are already moderately or highly excited, the true source for a planet's ejection --- secular evolution or the next external perturbation --- cannot be clearly identified. The mentioned fraction for the delayed ejection should therefore be treated with caution. We often see a strong planet--planet interaction subsequent to an encounter which leaves the system in a highly vulnerable state. It then only requires a very weak perturbation by another star to eject some of the planets which would not have been strong enough to disrupt the planetary system without the previous excitation. Those events are counted as delayed ejection even though they result from the combined effect of secular evolution and (another) prompt ejection due to the next encounter.

\begin{table}
    \centering
    \caption{Fractions of prompt and delayed ejections in the standard configuration of the 16k cluster for the different planet types.}
    \begin{tabular}{l|cc}
        \hline
        Planet & Prompt ejection & Delayed ejection \\
        \hline
        Jupiter & 88\% & 12\% \\
        Saturn & 61\% & 39\% \\
        Uranus & 52\% & 48\% \\
        Neptune & 59\% & 41\% \\
        \hline
    \end{tabular}
    \label{tab:prompt_vs_delayed_ejection}
\end{table}

\newKR{\cite{Fujii2019} perform $N$-body simulations of different cluster types and use a semi-analytical approach for the calculation of the fraction of ejected planets. The cluster model which is closest to one of our clusters is a high-density King-model cluster \citep{King1966} with $N = 2048$ and $W_0=3$. Using the power-law function from equation~10 in \cite{Fujii2019} and the corresponding best-fitting parameter for G-type stars, one obtains survival fractions [$1-f_\mathrm{ejc}(a)$] for the standard configuration of 93\% (Jupiter), 90\% (Saturn), 82\% (Uranus), and 76\% (Neptune). These values are higher than the results for our smallest cluster. The important difference between our work and \cite{Fujii2019} is not so much the different cluster models but the fact that we also take into account delayed ejections due to planet--planet scattering (for which the multiplicity of our planetary systems plays a crucial role) and the possibility of several strong encounters.}

\newKR{In the standard configuration of our 16k cluster the fraction of systems in which at least one planet is immediately ejected after an encounter (regardless of the intruder's mass) is 26$\%$. This value is comparable to the study of \cite{Malmberg2011} where they find fractions between 15 and 31$\%$ for a mass range of $0.6-1.5$\,M$_\odot$ for the intruder star. \cite{Malmberg2011} also determine the fractions of systems from which at least one planet has been ejected within 100\,Myr after the encounter and find fractions of 47--69$\%$ for flybys of stars with masses of $0.6-1.5$\,M$_\odot$. We find a corresponding value of 42$\%$ in the standard configuration of the 16k cluster (see Tab.~\ref{tab:fractions_of_1_ejection} for other configurations and other cluster sizes). The lower value likely stems from the shorter integration time. Although we simulate the planetary systems for 100\,Myr, the remaining simulation time after the first strong encounter is shorter wherefore our values cannot be directly compared with those from \cite{Malmberg2011}.}

\begin{table}
    \centering
    \caption{\KR{Fractions of planetary systems in which at least one planet is ejected during the simulation.}}
    \begin{tabular}{l|cccc}
        \hline
         \KR{Configuration}& \KR{8k} & \KR{16k} & \KR{32k} & \KR{64k}  \\
         \hline
         \KR{Standard} & \KR{34\%} & \KR{42\%} & \KR{42\%} & \KR{62\%}\\
         \KR{Compact} & \KR{34\%} & \KR{38\%} & \KR{38\%} & \KR{63\%}\\
         \KR{Resonant} & \KR{100\%} & \KR{100\%} & \KR{100\%} & \KR{100\%}\\
         \KR{Eccentric \#1} & \KR{35\%} & \KR{44\%} & \KR{43\%} & \KR{67\%}\\
         \KR{Eccentric \#2} & \KR{83\%} & \KR{82\%} & \KR{83\%} & \KR{90\%}\\
         \KR{Massive} & \KR{40\%} & \KR{52\%} & \KR{60\%} & \KR{77\%}\\
         \hline
    \end{tabular}
    \label{tab:fractions_of_1_ejection}
\end{table}

\KR{Although \cite{Malmberg2007} use $N$-body simulations, a direct comparison is also difficult as they only study encounters between stars but do not explicitly analyze planetary systems. Furthermore, their studied clusters are rather small in terms of stellar members. They define a star that has never been part of a binary system or has never undergone any close encounters with other stars as ``singleton''. We calculate the fraction of singletons in our simulations and obtain values of 50\% (8k cluster), 32\% (16k cluster), 20\% (32k cluster), and 6\% (64k cluster). \cite{Malmberg2007} provide the fraction of singletons for different half-mass radii and different numbers of cluster members. Taking their largest cluster ($N = 1000$) as reference, our results are most similar to the range of initial half-mass radii of 0.38--1.69\,pc.}

\SPZ{The most consistent simulations of star clusters with planetary systems so far have been performed by \cite{vanElteren2019}. They adopted the initial conditions from earlier simulations that tried to match the mass and size distributions of circumstellar discs in the Orion Trapezium cluster \citep{PortegiesZwart2016}. In the follow-up calculations by \cite{vanElteren2019}, the discs were replaced by planetary systems, selected according to the Oligarchic growth model \citep{Kokubo1998}. The parameter search was limited to a cluster of 1500 stars. The initial conditions were generated from the simulation of a star cluster with circumstellar discs of 400\,au each for 1\,Myr during which the discs were truncated and harassed by passing stars. In that time frame, the cluster evolved and the discs were affected by passing stars but not by internal processes. After 1\,Myr, a total of 977 stars remained bound in the cluster, 512 of which received a planetary system. The calculation was performed with 2522 planets with a total mass of 3527 Jovian masses. At an age of 11\,Myr, 10\,Myr after the birth of the cluster, 2165 planets were still bound to their host star: 16.5\% of the planets became unbound.
The majority ($\sim80$\,\%) of the ejected planets promptly escaped the cluster, the rest lingers around for at least half a million years before escaping the cluster potential.}

\subsection{Distribution in \texorpdfstring{$a$-$e$}{a--e} space}

\begin{figure*}
	\includegraphics[width=2\columnwidth]{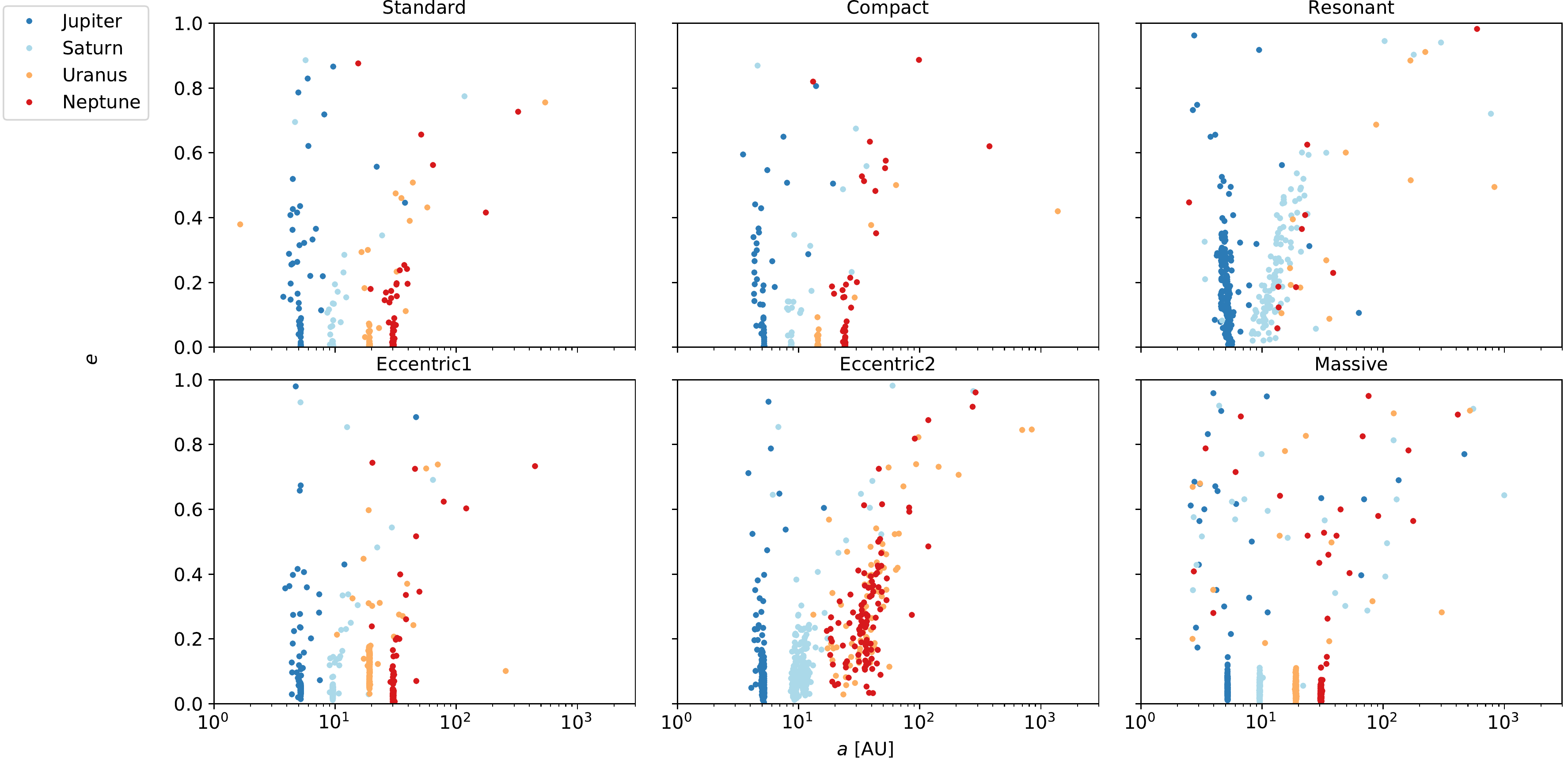}
    \caption{The $a$-$e$ space for all planets in the 8k star cluster which are not ejected from their host planetary system at $t=100$\,Myr, for the six different initial configurations. A video showing the $a$-$e$ space for the course of the simulation is available on our Silkroad project team webpage: \url{http://silkroad.bao.ac.cn/silkroad-save/a_e_space_N8k.mp4}.}
    \label{fig:a-e_space_8k}
\end{figure*}

We plot the eccentricity as a function of the semimajor axis of the planets for the different cluster sizes in Fig.~\ref{fig:a-e_space_8k} for the 8k cluster, and in Figs~\ref{fig:a-e_space_16k}, \ref{fig:a-e_space_32k}, and ~\ref{fig:a-e_space_64k} in the Appendix for the 16k, 32k, and 64k clusters, respectively. These figures clearly show a trend with increasing cluster size. In the 8k cluster most planets are only excited in eccentricity and just a few of them migrate into wider (or sometimes tighter) orbits. The fraction of highly eccentric orbits and planets that undergo significant orbital migration increases with increasing cluster density. The planets' distribution in the $a$-$e$ space is therefore wider for our larger clusters.

Having a look on the different initial orbital configurations reveals large differences in the $a$-$e$ space at the end of our simulations. In the 8k cluster (see Fig.~\ref{fig:a-e_space_8k}), the standard and compact configuration look similar. Most planets roughly retain their initial semimajor axis for 100\,Myr. Only  a few have migrated to larger semimajor axes and even less to tighter orbits (mainly Jovian-like planets, but there is also one Uranus-like planet in the standard case with $a<2$\,au). Especially the outermost planets tend to migrate to very wide orbits of more than 100\,au. In the standard configuration of the 8k cluster, even one Saturn-analogue can be found beyond $a>100$\,au. Due to the initially smaller semimajor axis of the outer three planets in the compact configuration, we observe a smaller number of planets on orbits with $a>50$\,au. In the compact configuration of the 16k cluster (see Fig.~\ref{fig:a-e_space_16k} in the Appendix) we can find in general more wide-orbit planets than in the 8k cluster but also wide-orbit planets with eccentricities $e<0.4$ which are missing in the standard configuration. The fraction of planets which remains unaffected in their orbital parameters is lower in the 32k and 64k clusters.

The distribution of planets in the $a$-$e$ space looks different for the resonant configuration. In the 8k cluster, most Jovian-like planets were at least excited in eccentricity and some also migrated within the system (mainly to wider orbits). None of the Saturn-like planets can retain its initial semimajor axis and eccentricity, and a clear trend towards wide, eccentric orbits is observable. The few Uranus-like planets which survived for 100\,Myr all have wide and/or eccentric orbits. Four of these planets have $a>100$\,au and one even has $a>800$\,au. All of these four planets have eccentricities of $e\ga 0.5$. While all Uranus-like planets failed to keep their initial semimajor axis, three of the Neptune-like planets succeeded in doing so. However, all of them were at least slightly excited in eccentricity (as well as the Uranus-like planets). In all of these three systems Uranus was ejected during the first few tens of thousands of years after a relatively short interaction with Neptune before the first strong encounter happened. Due to the encounters with Neptune, all three Uranus's migrated to an orbit with a semimajor axis smaller than that of Jupiter which led to the ejection of Uranus within the subsequent tens of thousands of years. This fortunate circumstance made the planetary system robust enough to withstand the gravitational perturbations by other stars during the remaining 99\,Myr. 

\begin{figure}
    \centering
    \includegraphics[width=0.47\textwidth, trim= 0.75cm 0.3cm 0.5cm 0.5cm,clip]{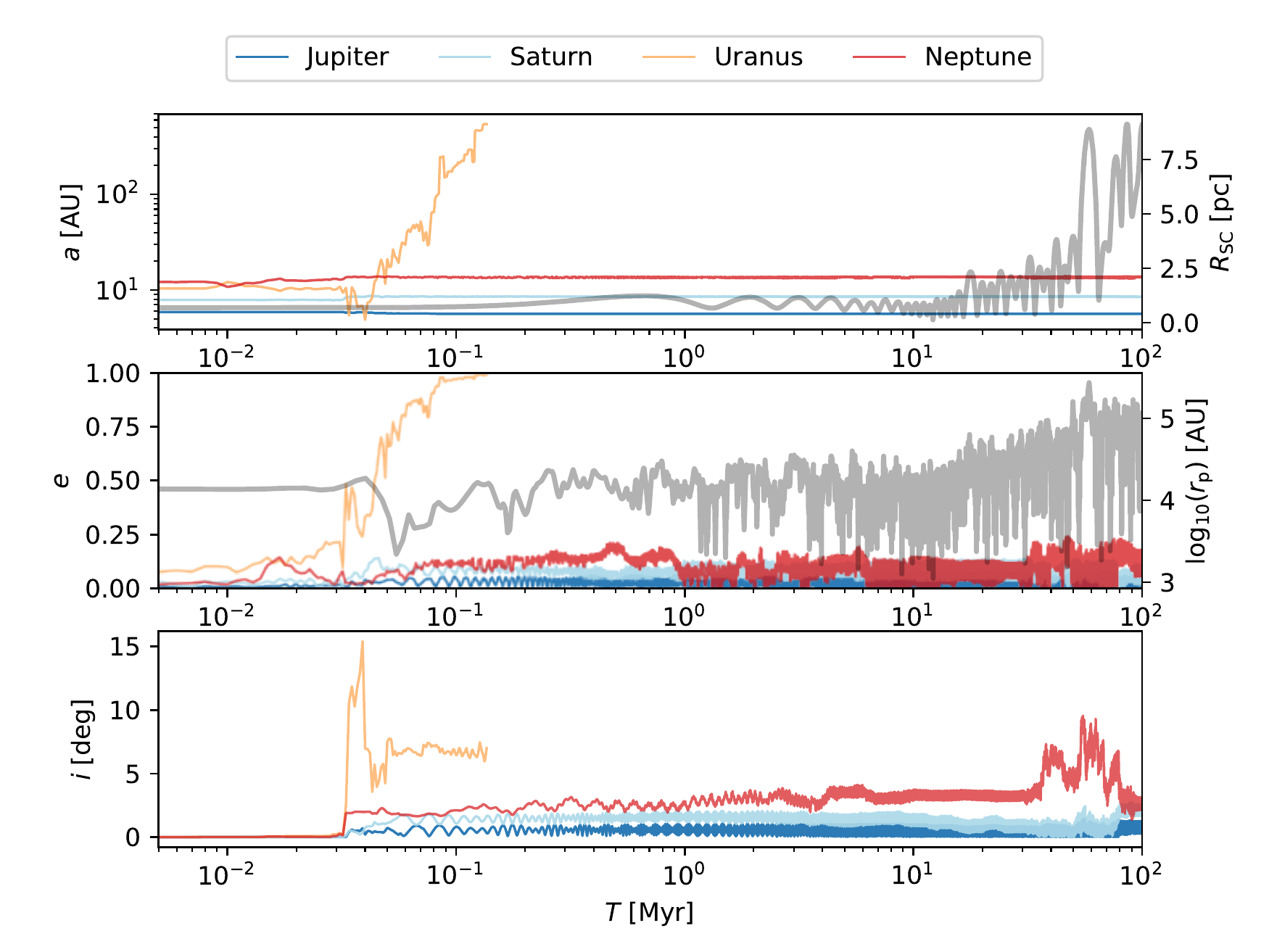}
    \caption{The orbital parameters $a$ (top), $e$ (middle), and $i$ (bottom) of a resonant planetary system from the 8k cluster as a function of time. The time is plotted in logarithmic scale. The scales on the right side correspond to the gray lines in the plots which represent the distance of the host star to the cluster centre (top) and the distance to the closest perturber (middle).
    }
    \label{fig:p_sys_194_N8k_resonant}
\end{figure}

The three orbital parameters $a$, $e$, and $i$ of one of these three planetary systems as a function of time are shown in Fig.~\ref{fig:p_sys_194_N8k_resonant}. The time is plotted in logarithmic scale to highlight the planet--planet scattering during the first 100,000 yr. We additionally plot the distance of the host star to the cluster centre and the distance to the next stellar perturber in gray in the top and middle panel to illustrate the interaction with the cluster. The cumulative gravitational effects of several neighbours in distances between 13\,000 and 23\,000\,au remove the resonance of Uranus and Neptune within the first 10\,000 yr which causes them to slightly interact with each other and to change their orbital position for a few hundred years. After migrating back to the second outermost position Uranus is already excited in eccentricity. The subsequent interaction with Jupiter and Saturn and the simultaneous close approach of a neighbouring star leads to the prompt ejection of Uranus and the removal of the remaining resonances in the system. Due to this circumstance, the remaining three planets form a stable system and stay relatively unperturbed for the rest of the simulation even though neighbouring stars closely approach the system several times.

The $a$-$e$ space for the first eccentric configuration in the 8k cluster looks similar to the standard and compact configuration but there are slight differences. On the one hand, the number of Jupiters that have high eccentricities is reduced. On the other hand, the number of Uranus- and Neptune-like planets with high eccentricities is larger in the first eccentric configuration. While Jupiter, Saturn, and Uranus all start at eccentricities of $e\approx 0.05$ there are Jupiters and Saturns that end up at nearly-circular orbits. However, this is not the case for Uranus. While some of them keep their initial eccentricity, there are no Uranus-like planets that have reduced it after 100\,Myr. Most of them are significantly excited in eccentricity.

\begin{figure}
    \centering
    \includegraphics[width=0.47\textwidth, trim= 0.75cm 0.3cm 0.5cm 0.5cm,clip]{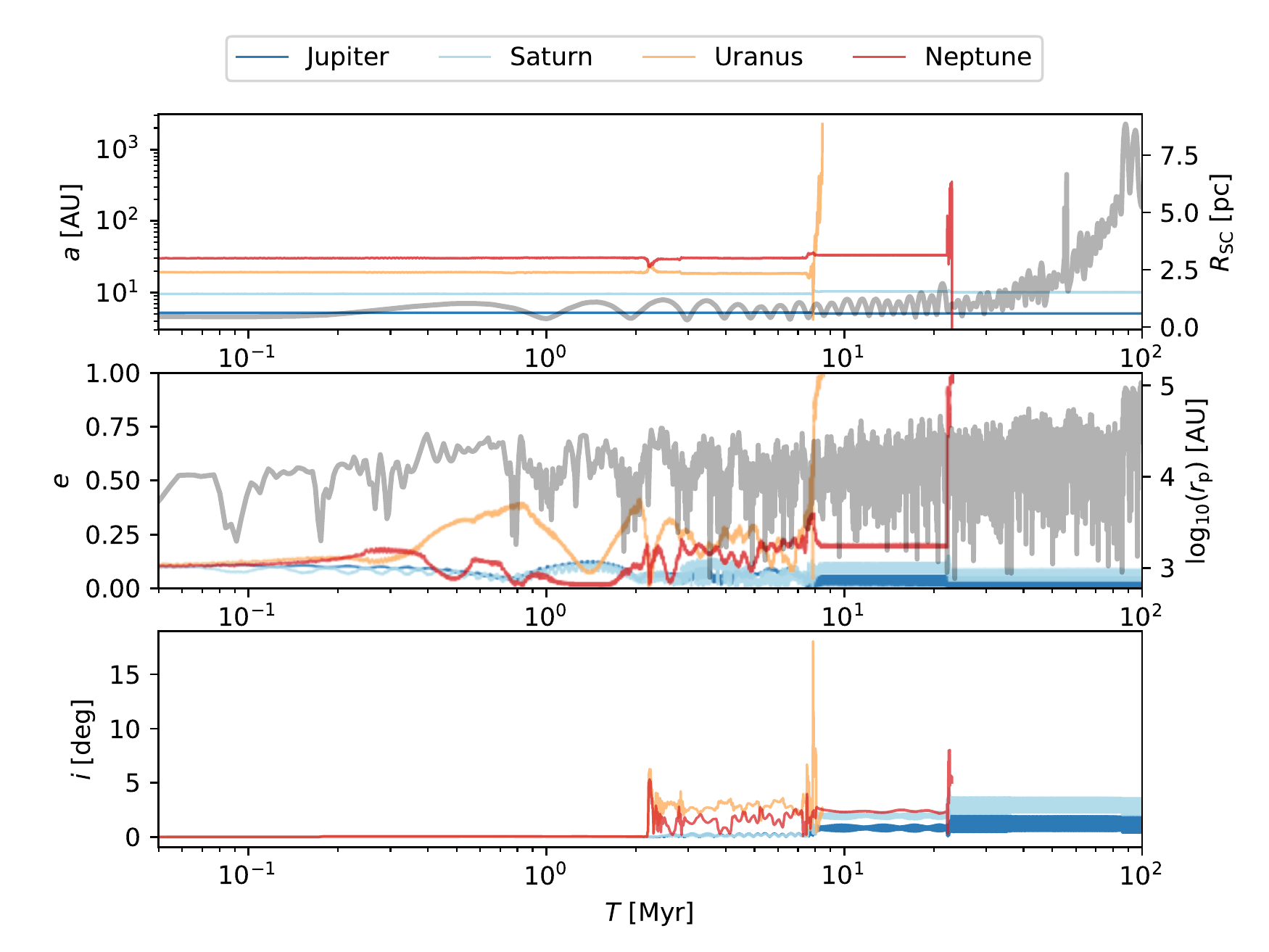}
    \caption{Same as in Fig.~\ref{fig:p_sys_194_N8k_resonant} but for a planetary system with initial eccentricities of $e=0.1$ from the 8k cluster.}
    \label{fig:p_sys_16_N8k_eccentric2}
\end{figure}

Increasing the initial eccentricity of all planets to $e=0.1$ makes a large difference in the outcome of our simulations. Especially Uranus and Neptune cover a much wider range in the $a$-$e$ space after 100\,Myr compared to the first eccentric configuration. On the other hand, most of our Jupiters and Saturns ``fall back'' to circular or almost circular orbits during the simulation which can be explained by the exchange of angular momentum between the planets during close encounters. This effect can be seen in Fig.~\ref{fig:p_sys_16_N8k_eccentric2} where Neptune is ejected at $t=22$\,Myr. A 6.9\,M$_\odot$ star approaches the planetary system down to a distance of 310\,au and ejects Neptune out of the system by ``kicking'' it to the inner regions of the planetary systems where it transfers angular momentum to Jupiter and Saturn. The eccentricity of both planets subsequently decreases.

\begin{figure}
    \centering
    \includegraphics[width=0.47\textwidth, trim= 0.75cm 0.3cm 0.5cm 0.5cm,clip]{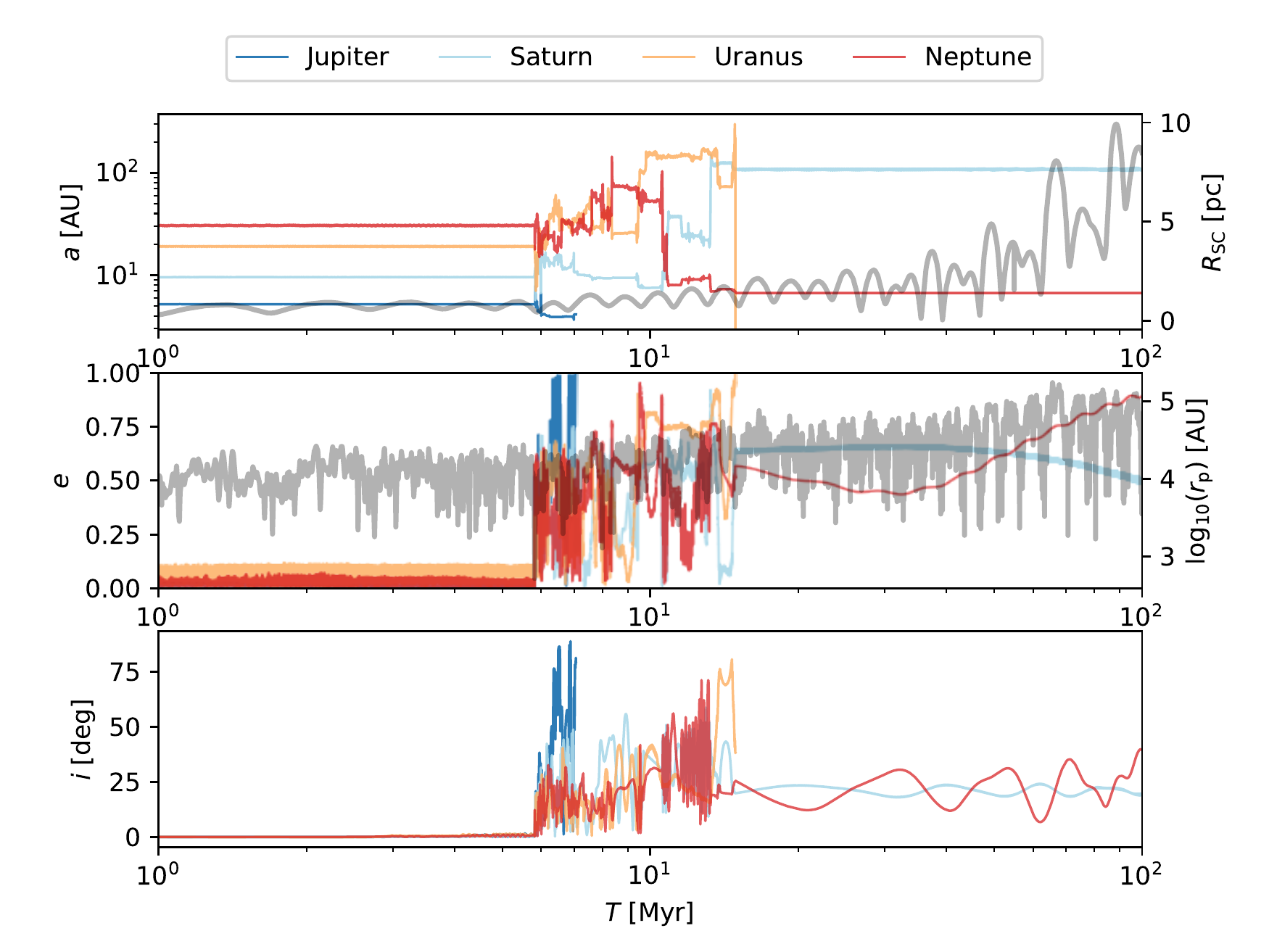}
    \caption{Same as in Fig.~\ref{fig:p_sys_194_N8k_resonant} but for a massive planetary system from the 8k cluster.}
    \label{fig:p_sys_0_N8k_massive}
\end{figure}

The massive configuration is more difficult to be excited in orbital parameters which can be seen the right bottom panel of Fig.~\ref{fig:a-e_space_8k}. There is a clear distinction between those planets that are only slightly perturbed, which are those with eccentricities below 0.2, and those which have been sufficiently perturbed to trigger fatal planet-planet scattering. Due to the equal mass of all planets in this configuration, the number of highly eccentric planets that have undergone orbital migration is almost comparable for all kind of planets. The numbers of almost unexcited planets is only slightly larger for the inner planets due to their smaller semimajor axis. Figure~\ref{fig:p_sys_0_N8k_massive} shows the orbital elements of a planetary system in which the planets are mostly unaffected for the first few million years despite several encounters. At $t=5.8$\,Myr a red dwarf with a mass of 0.3\,M$_\odot$ approaches the system closer than 240\,au causing a transfer of energy and angular momentum from Neptune to the perturber. Due to the inwards migration of Neptune on an orbit with an eccentricity of around 0.5, a fatal chain reaction with strong planet--planet scattering is triggered in which the eccentricity of the three other planets is excited as well. After a very short change of position with Saturn, Jupiter migrates inwards and reaches twice an eccentricity of more than 0.9 before it is finally ejected at $t=7.1$\,Myr. During that time and in the following 7 Myr the remaining planets Saturn, Uranus, and Neptune change their order several times. Due to that strong interaction, Uranus migrates outwards to a very wide and eccentric orbit. Saturn follows at $t=13.3$\,Myr which finally leads to the ejection of Uranus at $t=14.9$\,Myr. For the remaining 85 Myr, Saturn retains its wide orbit of $a\approx 107$\,au while Neptune remains at $a=6.7$\,au.

\subsection{Distribution in \texorpdfstring{$a$-$i$}{a--i} space}

\begin{figure*}
    \centering
    \includegraphics[width=2\columnwidth]{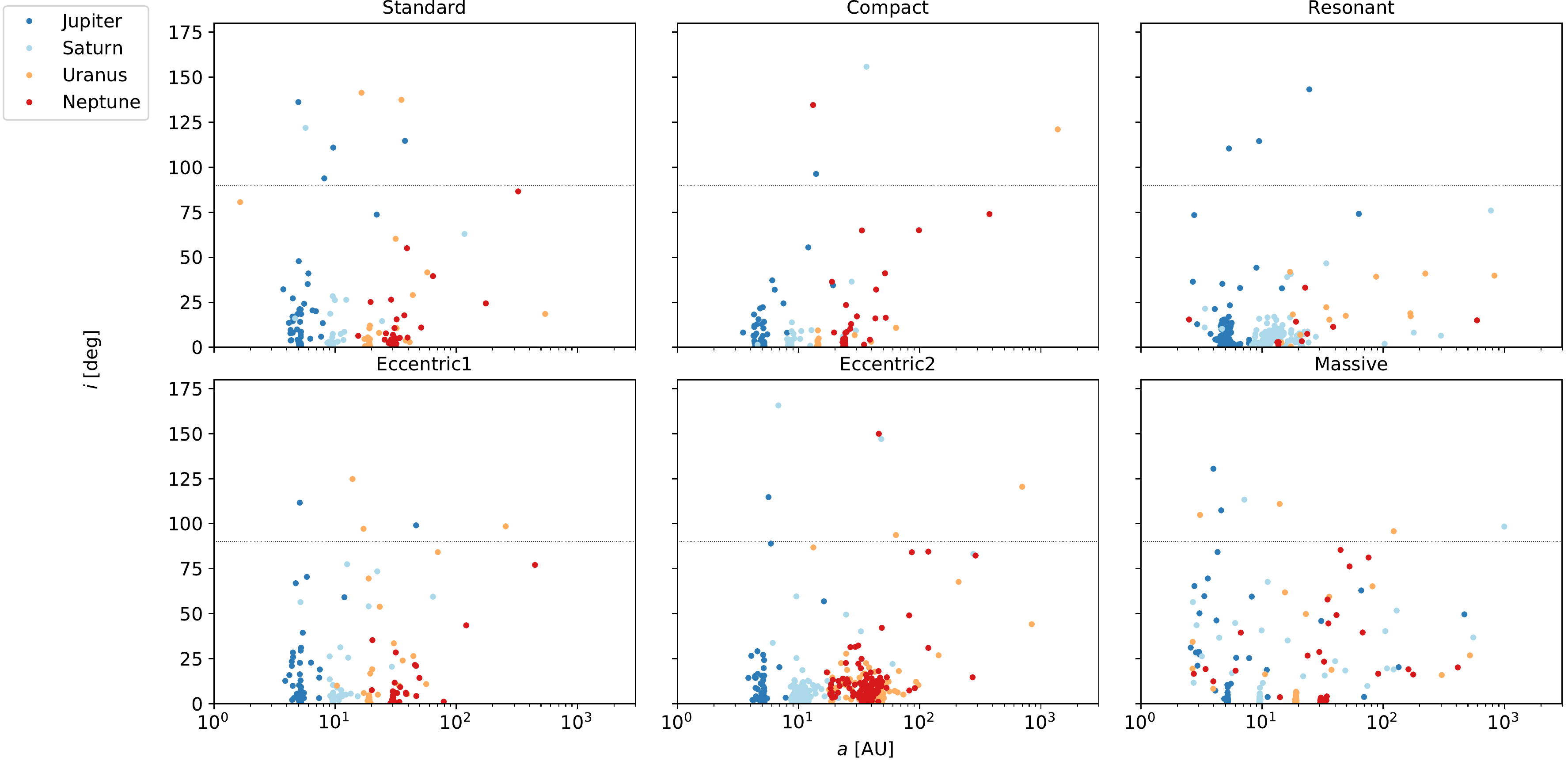}
    \caption{The $a$-$i$ space for all planets in the 8k star cluster which are not ejected from their host planetary system after a simulation time of 100\,Myr for the six different initial configurations. The dotted black line shows the threshold of $i=90^\circ$. Planets near that value have polar orbits while those above it have retrograde orbits.}
    \label{fig:a-i_space_N8k}
\end{figure*}

A change in eccentricity is often directly related to a change in inclination since both result from the transfer of angular momentum. By looking on the $a$-$i$ space of the planets after 100\,Myr for the different cluster sizes in Figs~\ref{fig:a-i_space_N8k}, \ref{fig:a-i_space_N16k}, \ref{fig:a-i_space_N32k}, and~\ref{fig:a-i_space_N64k}, we can again see a wide distribution in that parameter space, although all planets started on coplanar, prograde orbits (Figs \ref{fig:e-i_space_N8k}, \ref{fig:e-i_space_N16k}, \ref{fig:e-i_space_N32k}, and~\ref{fig:e-i_space_N64k} show the $e$-$i$ space after 100\,Myr for comparison). Those planets that get excited to polar orbits ($i \approx 90^\circ$) or retrograde orbits ($i > 90^\circ$) are of special interest.

In 2006, \cite{Remijan2006} found first evidence that parts of the protoplanetary disc around the binary system IRAS 16293–2422 are counterrotating which means that planets that form in that region would have a retrograde orbit. The first two detected planets for which a retrograde or polar orbit is assumed are WASP-17b \citep{Anderson2010} and HAT-P-7b \citep{Winn2009}.

\begin{figure}
    \centering
    \includegraphics[width=0.47\textwidth, trim= 0.75cm 0.3cm 0.5cm 0.5cm,clip]{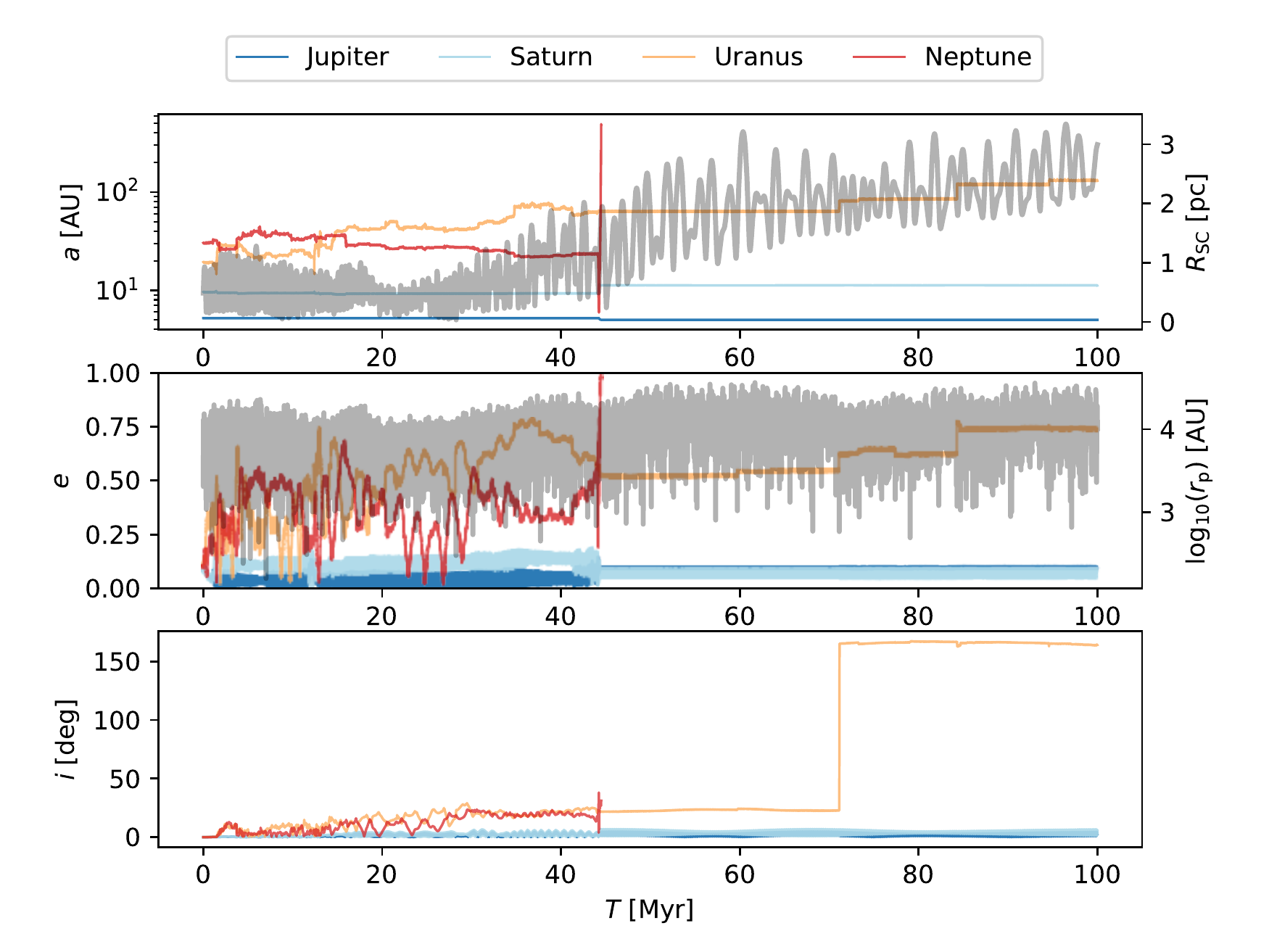}
    \caption{Same as in Fig.~\ref{fig:p_sys_194_N8k_resonant} but for a planetary system with initial eccentricities of $e=0.1$ from the 32k cluster. The time is plotted in linear scale.}
    \label{fig:p_sys_74_N32k_eccentric2}
\end{figure}

We find planets with inclined orbits of more than $90^\circ$ in all of our simulations, independent of the initial configuration and stellar density of the host cluster. Some planets switch to a retrograde orbit for only a few million years but some also keep their highly inclined orbit for the rest of the simulation. An example of the latter case is shown in Fig.~\ref{fig:p_sys_74_N32k_eccentric2}. At 71\,Myr, the encounter of a  1.9\,M$_\odot$ star of less than 600\,au causes Uranus (the outermost planet at that time) to switch from a prograde orbit (inclined by $23^\circ$) to a retrograde orbit with $i=164^\circ$. Despite several additional encounters during the remaining 29\,Myr with periastron distances of less than 1000\,au, Uranus keeps its retrograde orbit until the end of the simulation.

There is no clear trend visible in which configuration or with which cluster density we can expect the highest fraction of retrograde orbits. However, in all four cluster simulations, the massive configuration results in the largest number of highly inclined orbits with $i>50^\circ$ after 100\,Myr. We can therefore conclude that retrograde orbits mainly occur due to strong external perturbation while planet--planet scattering especially seems to be an additional source for the excitation of planetary orbits to the range of $i\approx 50^\circ-80^\circ$.

\subsection{Dynamical evolution of a planetary system in different initial configurations}

\begin{figure*}
    \centering
    \includegraphics[width=0.49\textwidth, trim= 0.2cm 0.0cm 0.2cm 0.0cm,clip]{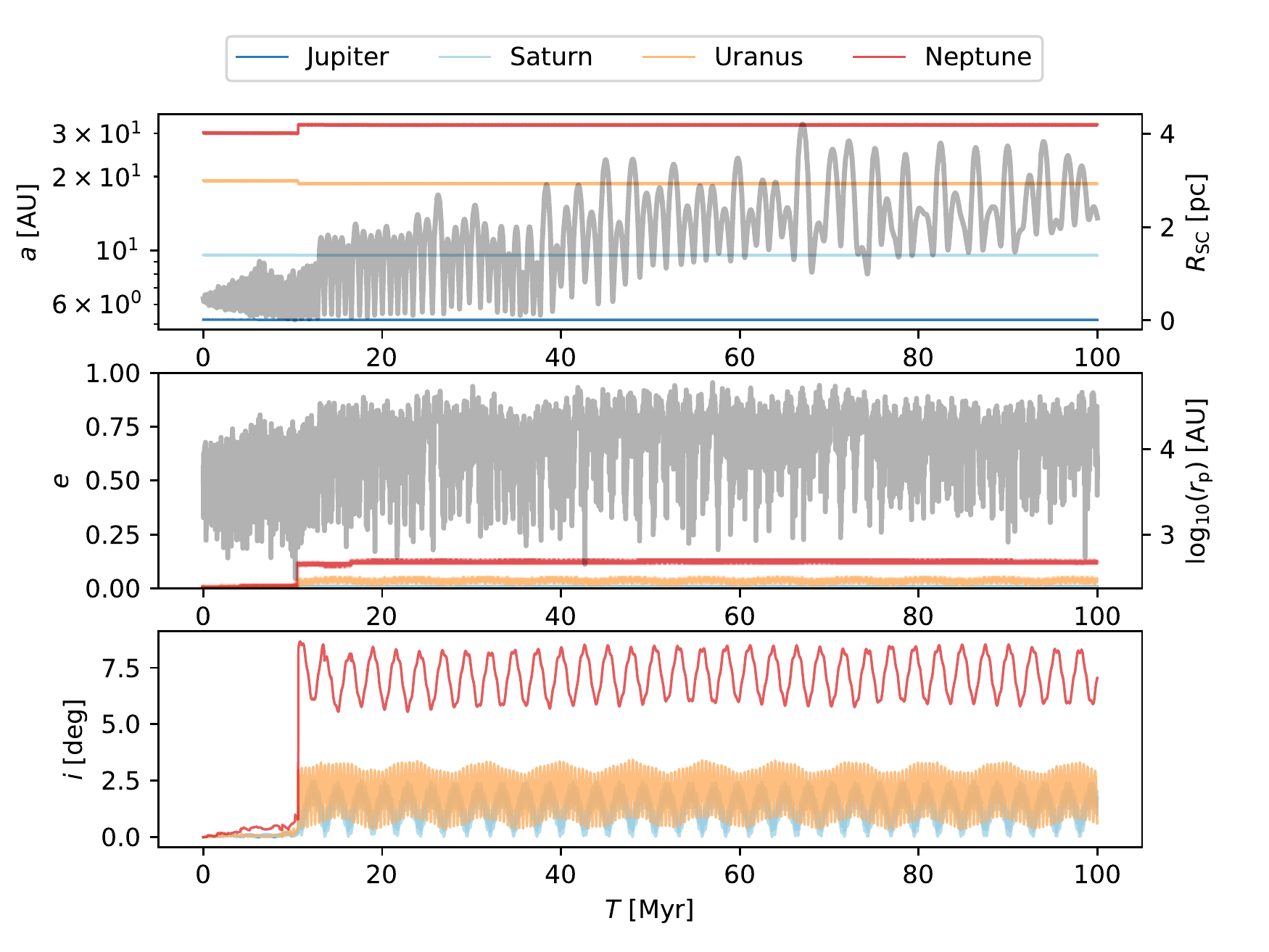}
    \includegraphics[width=0.49\textwidth, trim= 0.2cm 0.0cm 0.2cm 0.0cm,clip]{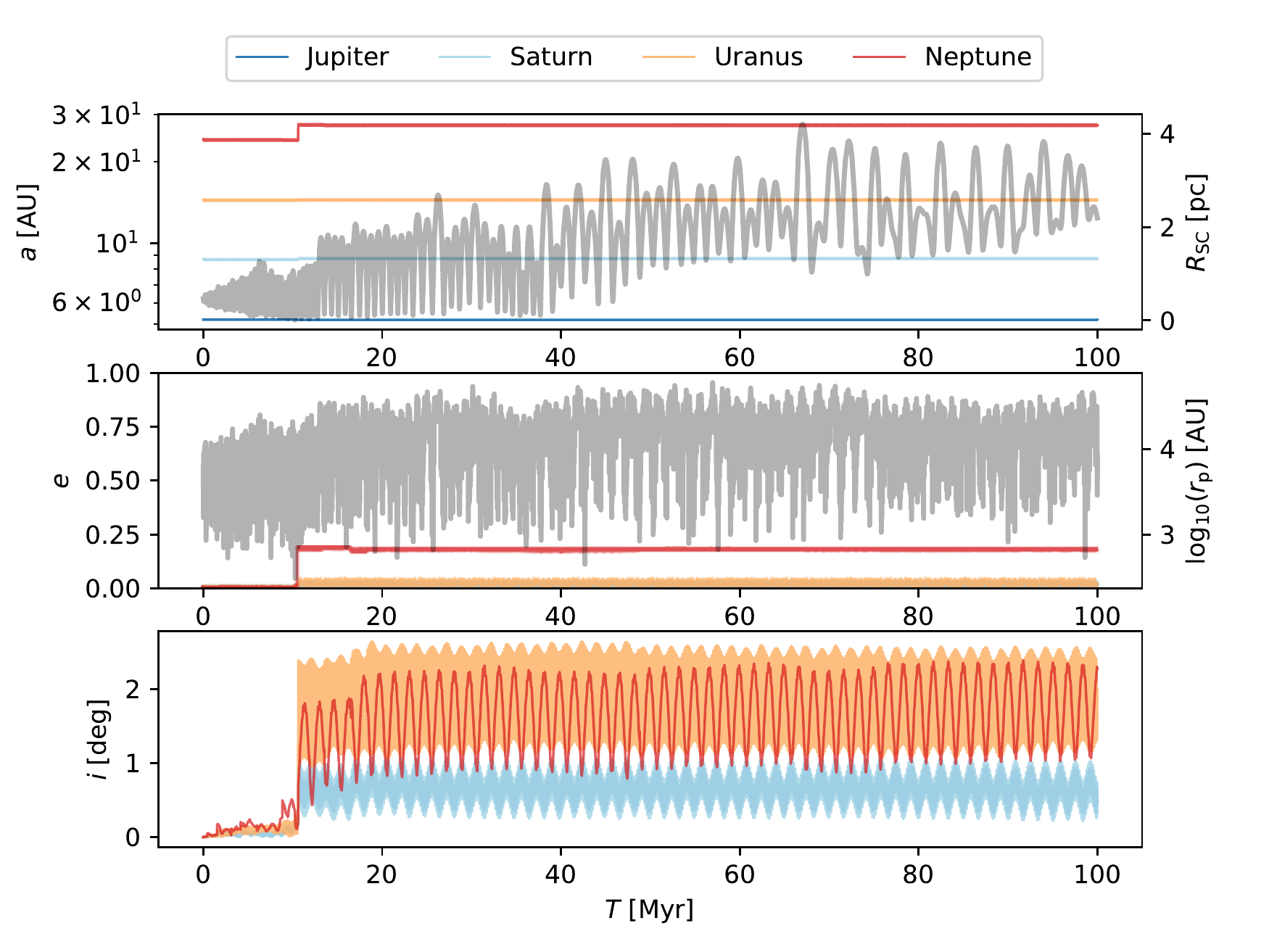}
    \includegraphics[width=0.49\textwidth, trim= 0.2cm 0.0cm 0.2cm 0.0cm,clip]{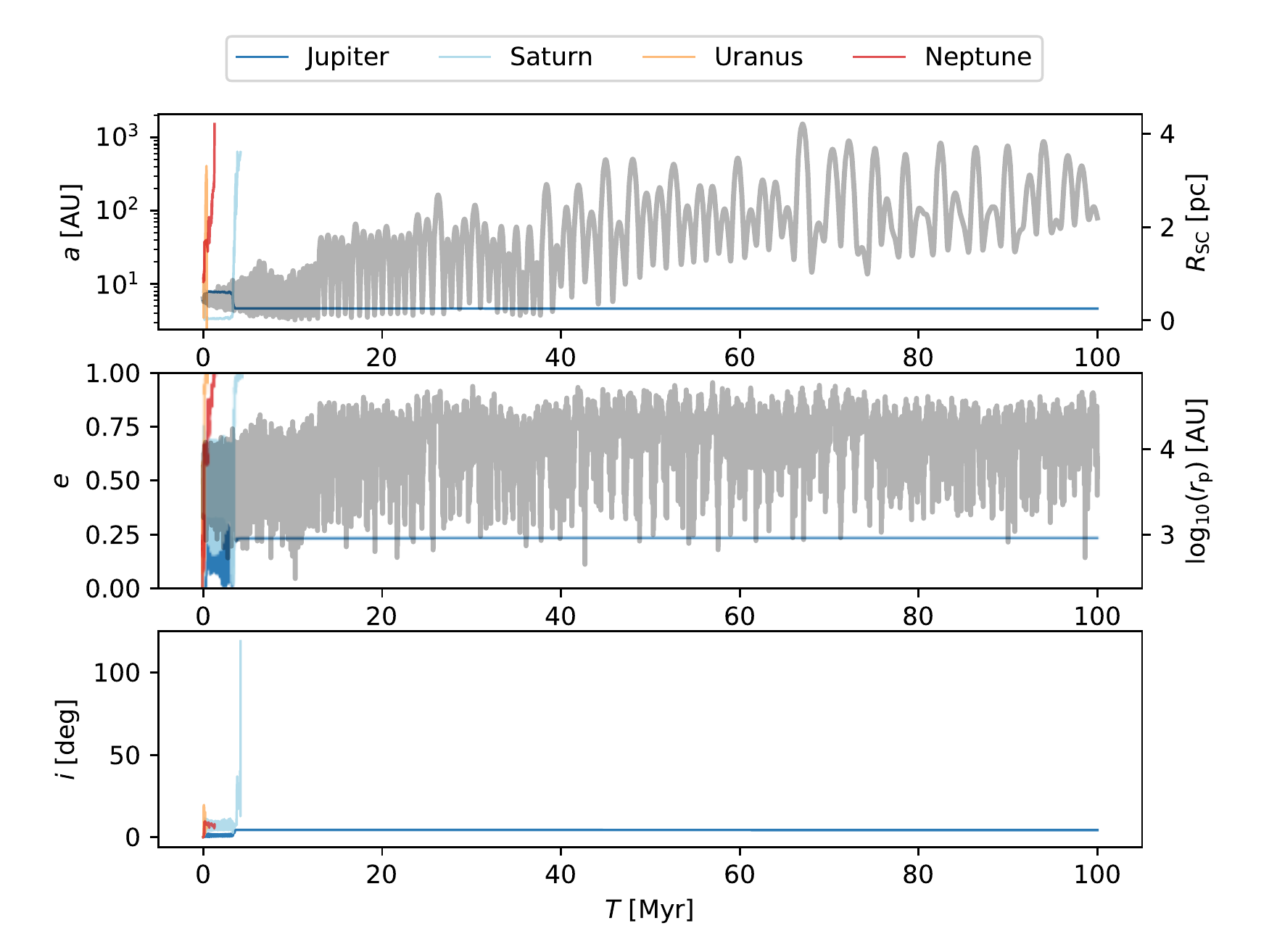}
    \includegraphics[width=0.49\textwidth, trim= 0.2cm 0.0cm 0.2cm 0.0cm,clip]{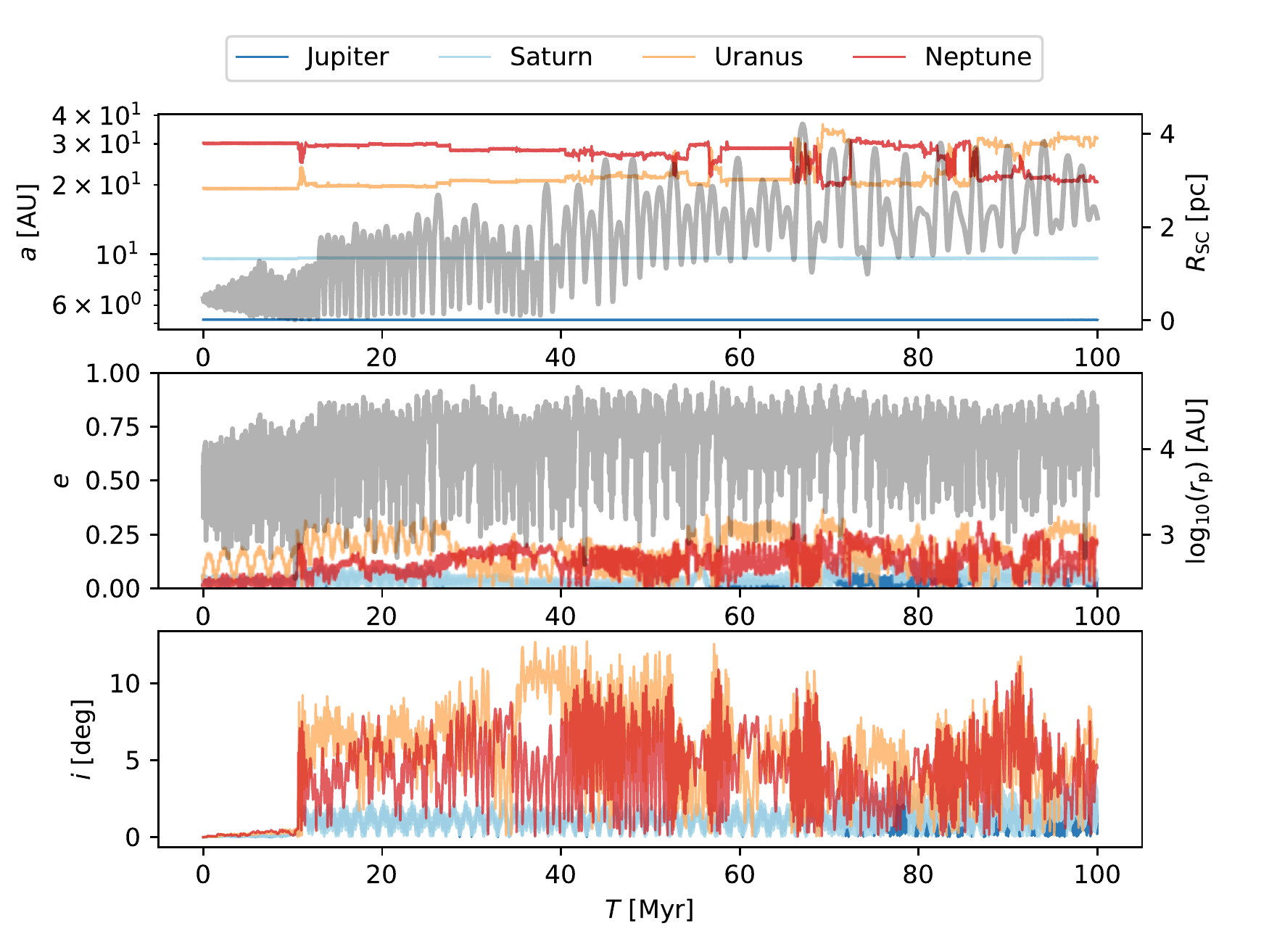}
    \includegraphics[width=0.49\textwidth, trim= 0.2cm 0.0cm 0.2cm 0.0cm,clip]{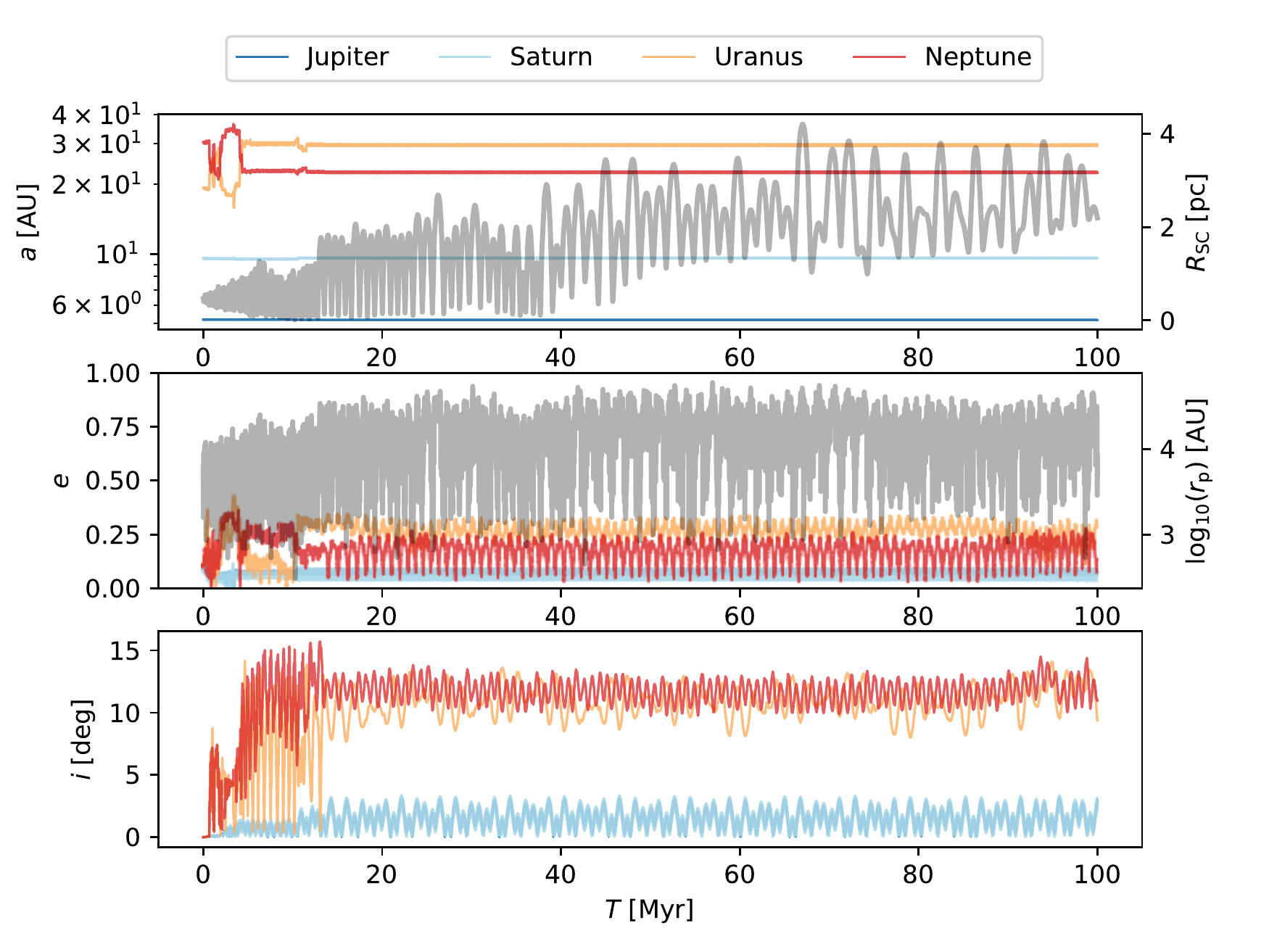}
    \includegraphics[width=0.49\textwidth, trim= 0.2cm 0.0cm 0.2cm 0.0cm,clip]{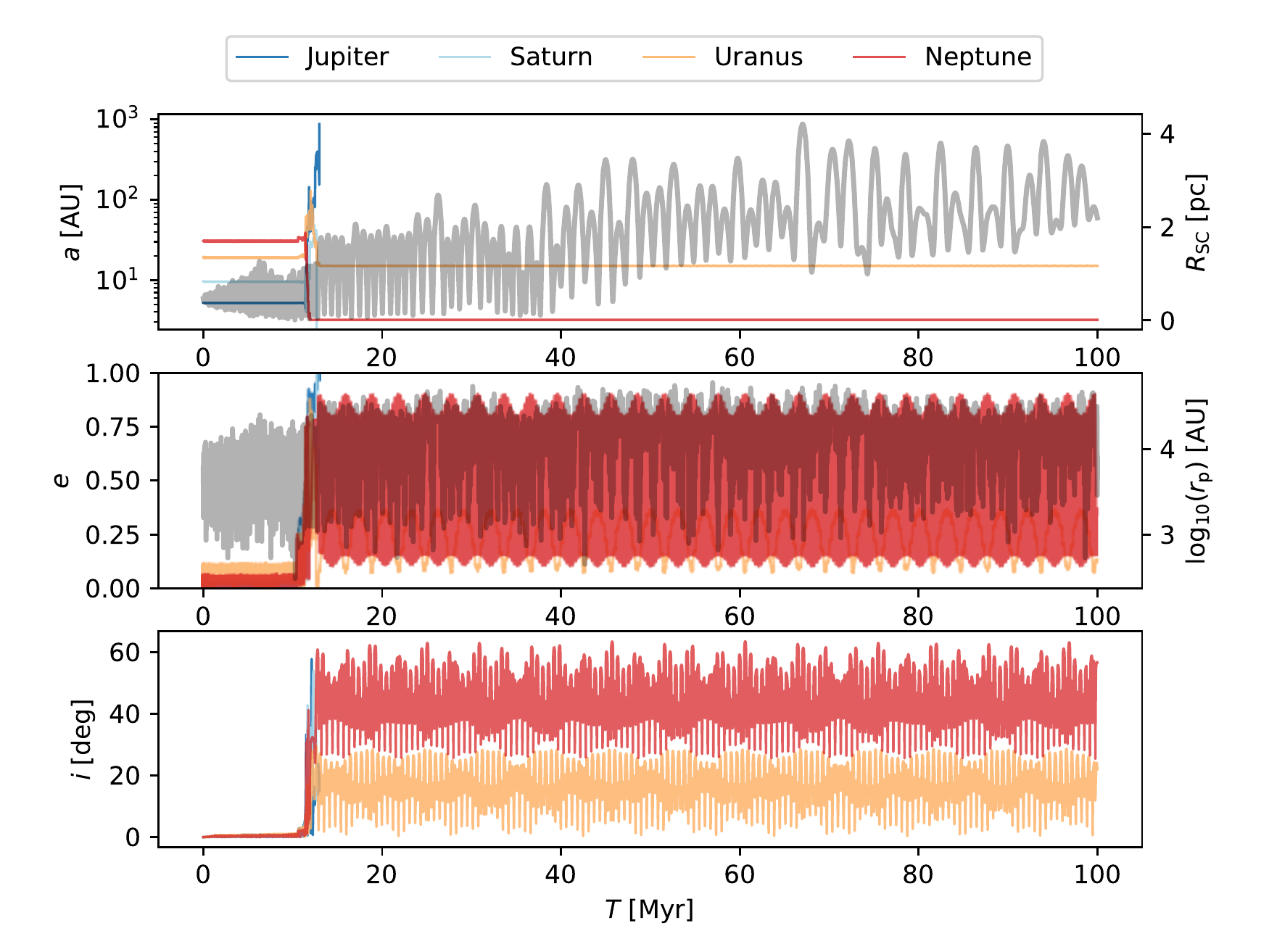}
    \caption{Comparison of the dynamical evolution of one planetary system in the 32k cluster around the same host star in all six different initial configurations. \textit{}{Top left:} Standard configuration. \textit{Top right:} Compact configuration. \textit{Middle left:} Resonant configuration. \textit{Middle right:} Eccentric\#1 configuration. \textit{Bottom left:} Eccentric\#2 configuration. \textit{Bottom right:} Massive configuration.}
    \label{fig:comparison_p_sys_15_N32k}
\end{figure*}

In the previous sections, we have shown the differences between the initial configurations averaged over identical 200 planetary systems. However, from this we can only have a rough estimate of how the dynamical evolution of one planetary system looks like if we put it in different initial configurations. Therefore, we show the dynamical evolution of planetary system \#15 from the 32k cluster in all six different configurations in  Fig.~\ref{fig:comparison_p_sys_15_N32k} as an example. While in the standard case Jupiter and Saturn only get slightly excited in eccentricity and inclination due to an encounter at $t=10$\,Myr, Uranus migrates inwards by $\sim 0.5$\,au and Neptune outwards by $\sim 2$\,au. Neptune's increase in eccentricity and inclination is the largest of all four planets.

There is almost no difference in the dynamical evolution of Jupiter, Saturn, and Uranus between the standard and compact configuration. However, instead of migrating inwards, Uranus keeps its semimajor axis in the compact configuration unlike Neptune which now migrates outwards by 3.5\,au due to the same encounter as in the standard configuration. Neptune is also more excited in eccentricity but less in inclination.

In the resonant configuration, Saturn, Uranus, and Neptune do not even survive until the first very close encounter at $t=10\,Myr$ which is the only encounter in the standard and compact configuration which affects the system significantly. Uranus and Neptune are both ejected within the first 1.3\,Myr, whereas Saturn migrates to $a=3.4$\,au and strongly oscillates in eccentricity within the following millions of years. Jupiter, which migrates outwards to a semimajor axis of 7.8\,au, and Saturn cross their orbits after an additional stellar encounter at $t=3.3$\,Myr which causes the ejection of Saturn at $t=4.2$\,Myr. Due to that interaction, Jupiter migrates back to a smaller orbit of $a=4.6$\,au and stays completely unaffected during additional encounters within the rest of the simulation.

The dynamical evolution of the first eccentric configuration is characterized by the interaction between Uranus and Neptune. The same encounter that affected the standard and compact configuration causes a first close orbital approach of Uranus and Neptune after roughly 10\,Myr. Due to that, their initially small eccentricities increase as well as their inclinations. Additional encounters, especially during the last 50 Myr of our simulation cause steady interaction and switch of orbital positions between the two outermost planets. Jupiter and Saturn stay relatively unaffected in this simulation.

The higher initial eccentricity in the second eccentric configuration leads to a quicker interaction between Uranus and Neptune than in the previous case. After a switch of positions the whole planetary system stays stable for the rest of the simulation. 

The massive configuration reveals the increasing risk for the innermost planets if all planets in the system have the same mass. In all of the previous configurations, Jupiter survived without facing any serious dangers for its orbital stability while Saturn survived in four configurations. In the massive configuration, these two planets are the only planets that get ejected while Uranus and Neptune survive. The external perturbation that leads to the ejection of Jupiter and Saturn is the same stellar encounter which is the formative encounter in the dynamical evolution of the system in the standard, compact and first eccentric configuration.

\section{Discussion and conclusions}\label{sec:conclusion}

We have explored the stability and vulnerability of 4800 planetary systems, which are exposed to repeated stellar encounters in the star cluster in which they formed. In each of our four star clusters, we distribute 200 identical planetary systems in six different initial configurations, which were inspired by the Monte Carlo simulations of \cite{Li2015}. All planetary systems were Solar system analogues (with host star masses of $\sim 1$\,M$_\odot$) consisting of the solar system's gas giant planets Jupiter, Saturn, Uranus, and Neptune. In the standard configuration, the planets have their current semimajor axes but circular orbits. Two other configurations are more compact versions of that case which are called the compact and resonant configurations (due to mutual MMRs between the planets). In two additional configurations, the eccentricities of the standard case are increased to the planets' present-day values and to larger values of $e=0.1$ (the first and second eccentric configurations). The sixth configuration differed from the standard configuration only in the equal planetary masses of one Jovian mass.

Our results for the cluster simulation can be summarized as follows: after 100\,Myr the star clusters have undergone the phases of mass segregation, stellar mass-loss and core collapse, and re-expand again after a time of maximum central density. The maximum central density reaches about 10 times the initial central density; after mass segregation and core collapse the cluster generally re-expands and reaches a quasi-stationary state, where the central density is about equal to its initial value, and the average density inside the half-mass radius has dropped by a factor of approximately 10. We observe that at that stage most of the dynamical interactions between planetary systems and passing stars are over, so for the current pilot study we stop our models at 100\,Myr. 

Generally, the most stable planetary systems are the standard and compact ones, and the configuration with small (current) eccentricities. The results for the standard, compact, and first eccentric configuration are comparable in fractions of surviving planets and final distribution in $a-e$ and $a-i$ space. However, a trend is observable that the compact system becomes slightly more resistant and the eccentric one slightly more vulnerable with increasing stellar density relative to the standard case.

We note that the compact system relative to the standard system shows very little differences --- one would expect that it experiences less strong interactions under the effect of the same encounters as the standard system; our result of very similar survival fractions can only be explained by stronger internal interactions, which destabilize the system even after relative weak perturbations. Furthermore, small initial eccentricities seem to not significantly change the vulnerability of a planetary system.

Due to its innermost position and highest mass, Jupiter is generally the planet with the highest chance to survive a perturbation by a stellar encounter of another cluster member, followed by Saturn. The exact order of the survival fraction of the two outermost planets Uranus and Neptune depends on the initial configuration and cluster density. However, usually Uranus is slightly more likely to be ejected from the planetary system due to an encounter or secular evolution. Even though Uranus is not the outermost planet, its lower mass makes the planet slightly more vulnerable to gravitational perturbations from the host cluster due to its lower gravitational binding energy compared to Neptune. This difference can especially be seen in the survival fractions for the second eccentric configurations. In all four clusters, Uranus has by far the lowest chance to survive in the system if all planets are started with their true semimajor axes but with an eccentricity of 0.1. If all planets have equal masses, the differences in survival fractions shrink significantly. Due to its smallest semimajor axis, Jupiter still has a slightly higher chance for survival while the rates for Saturn, Uranus, and Neptune are almost equal. From this, we can deduce that a planet's mass (compared to the other planets in the system) plays a more crucial role for the estimation of its vulnerability than its semi-major axis.

The fourth most stable system is the massive configuration in the 8k and 16k cluster but the system with initial eccentricities of $e=0.1$ is instead more resistant in the 32k and 64k cluster. In all clusters, the resonant system is the one with the highest vulnerability. However, the system is a special case and very interesting for a certain reason. Our integrations show that without perturbations by passing stars it is generally very short lived, getting unstable after about $10^5$ yr, around that time Uranus and Neptune are inevitably ejected from the planetary system. However, embedded in a star cluster, the system tends to be more stable. We believe that this is due to a process where stellar encounters detune or break the resonances and thus render the systems more stable. In many of the simulated systems, this is achieved by only ejecting one of the outer planets (Uranus or Neptune), and then the remaining three-planet system survives much longer than in the isolated case.

In \cite{vanElteren2019}, the authors find that the probability of a star to lose a planet is independent of the planet mass and independent of its initial orbital separation. As a consequence, the mass distribution of free-floating planets 
would be indistinguishable from the mass distribution of planets bound to their 
host star. Our results do not confirm this.
\SPZ{The discrepancy may result from the larger number of stars in the clusters in our simulations, the longer evolutionary time-scales (we integrated for 100\,Myr whereas 
in \cite{vanElteren2019} they integrated up to 10\,Myr), and finally they adopted the Oligarchic growth model for planetary systems. In the latter model, planet mass increases further away from the host star. This has interesting consequences for the stability of the planetary systems from perturbations from inside as well as for external perturbations. A small perturbation from another star may render an entire planetary system catastrophically unstable, whereas if the outer most planets have low mass, such a system survives more easily in a dense stellar environment.}

The survival fractions for the different planet types in our simulations are generally smaller than those of \cite{Li2015}. This is due to the different approaches. First, \cite{Li2015} randomly select their encounter parameter equally from the available phase space that is not realistic \citep[see figs 1 and 2 in][]{Spurzem2009}. Secondly, \cite{Li2015} only focus on the prompt ejections of planets while we continue the integration of the planetary systems long enough to account for secular evolution. Thirdly, our planetary systems are exposed to the cumulative effect of several encounters over a significant fraction of the host star cluster's lifetime. From the reduced survivability of the planets, which we see in our results compared to \cite{Li2015}, we can conclude that the effects of secular evolution and cumulative encounters are not negligible.

We find that passing stars excite mutual inclinations between planets in our planetary systems; quite some cases lead to high values of relative inclination and even to counter-rotating planets. It is quite impossible to excite significant inclinations by internal evolution of planetary systems, they are a tell-tale sign of the important role of stellar encounters in shaping the planetary system. While this effect has been mentioned in previous studies \citep[such as in][]{Spurzem2009}, there is not yet a more quantitative study of this process.

Our simulations could be and will be refined in future work in many ways. Planetary systems around more massive stars are subject to orbital changes when the host star becomes a red giant and finally loses significant mass. The presence of many initial binaries, which is expected from star and cluster formation, will be an interesting issue ---  including S- and P-type planetary systems.

The Monte Carlo models of \cite{Li2015} give some information about encounters between planetary systems and binary stars. Finally, in this work we have only presented a limited set of star clusters. A wider parameter study may be required to predict the impact of stellar encounters on the final planetary population in the Galactic field. Other processes shaping planetary systems in the formation process inside a star cluster have also not been taken into account here. 

We have, however, clearly shown that encounters of passing stars in star clusters have a considerable effect and contribute to the diversity of planetary systems in all respects.

\section*{Acknowledgements}
We want to thank Rosemary Mardling, Willy Kley, and Gongjie Li for useful discussions.
 KS and RS acknowledge the support of the DFG priority program SPP 1992 ``Exploring the Diversity of Extrasolar Planets''
  (SP~345/20-1). RS has been supported by National Astronomical Observatories of Chinese Academy of Sciences, Silk Road Project, and by National Natural Science Foundation of China under grant no. 11673032. All simulations have been done on the GPU accelerated cluster ``kepler'', funded by Volkswagen Foundation grants 84678/84680.
MBNK was supported by the National Natural Science Foundation of China (grant 11573004) and by the Research Development Fund (grant RDF-16-01-16) of Xi'an Jiaotong-Liverpool University (XJTLU).

\section*{Data availability}
The data underlying this article will be shared on reasonable request to the corresponding author.




\onecolumn
\appendix

\section{Additional Material}

\begin{figure*}
	\includegraphics[width=1\columnwidth]{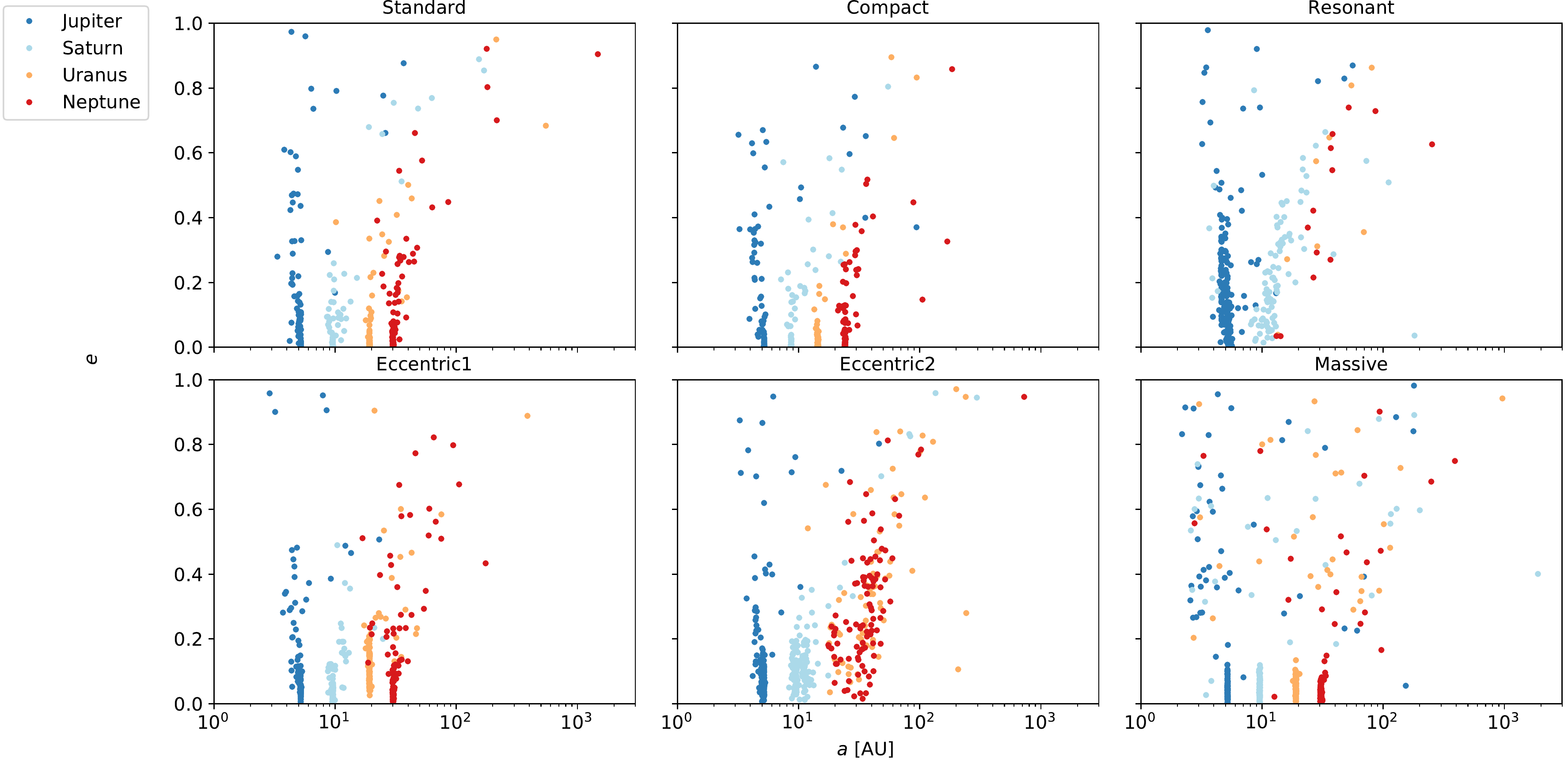}
    \caption{Same as Fig.~\ref{fig:a-e_space_8k} but for the 16k cluster.}
    \label{fig:a-e_space_16k}
\end{figure*}

\begin{figure*}
	\includegraphics[width=1\columnwidth]{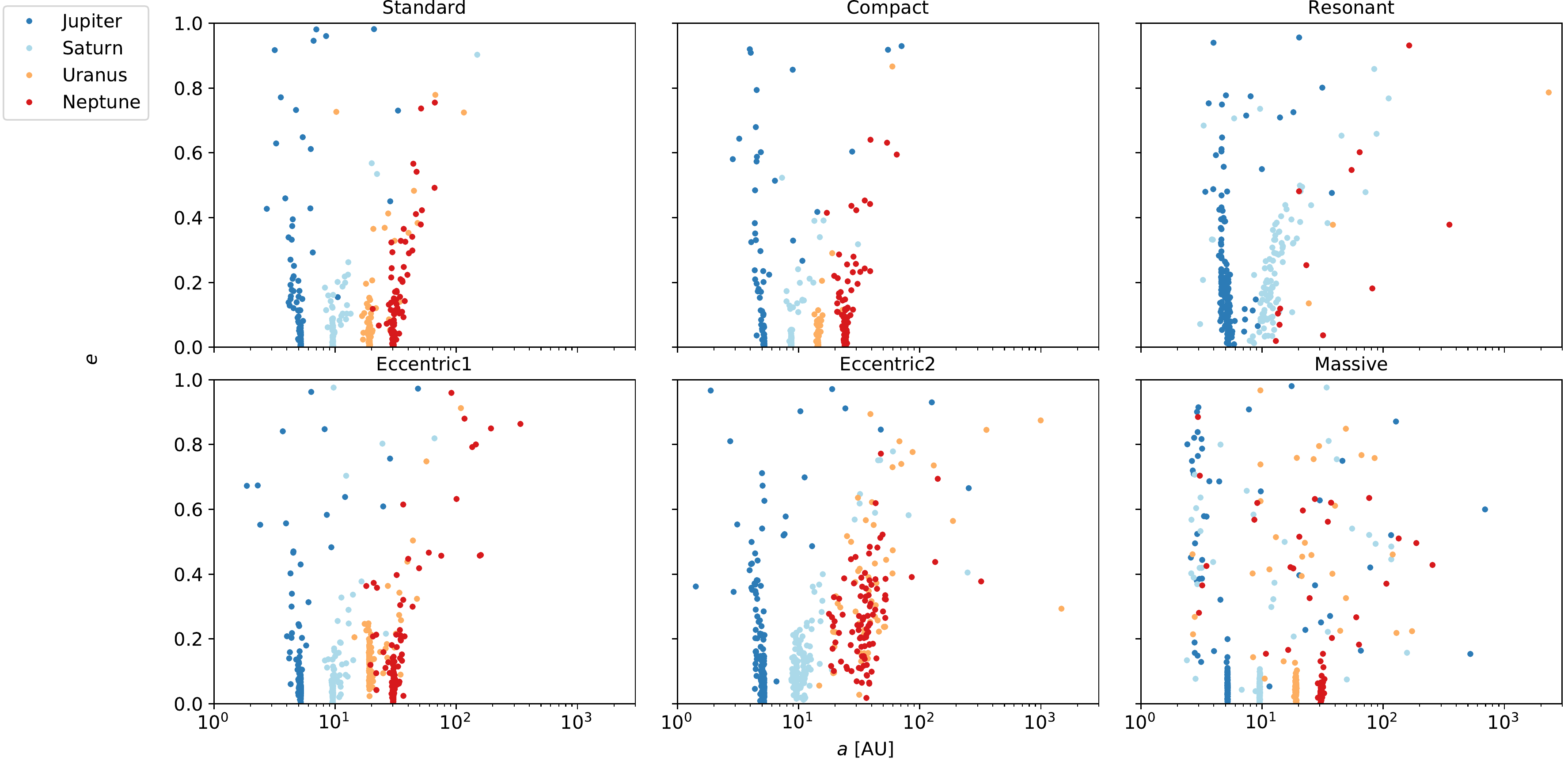}
    \caption{Same as Fig.~\ref{fig:a-e_space_8k} but for the 32k cluster.}
    \label{fig:a-e_space_32k}
\end{figure*}

\begin{figure*}
	\includegraphics[width=1\columnwidth]{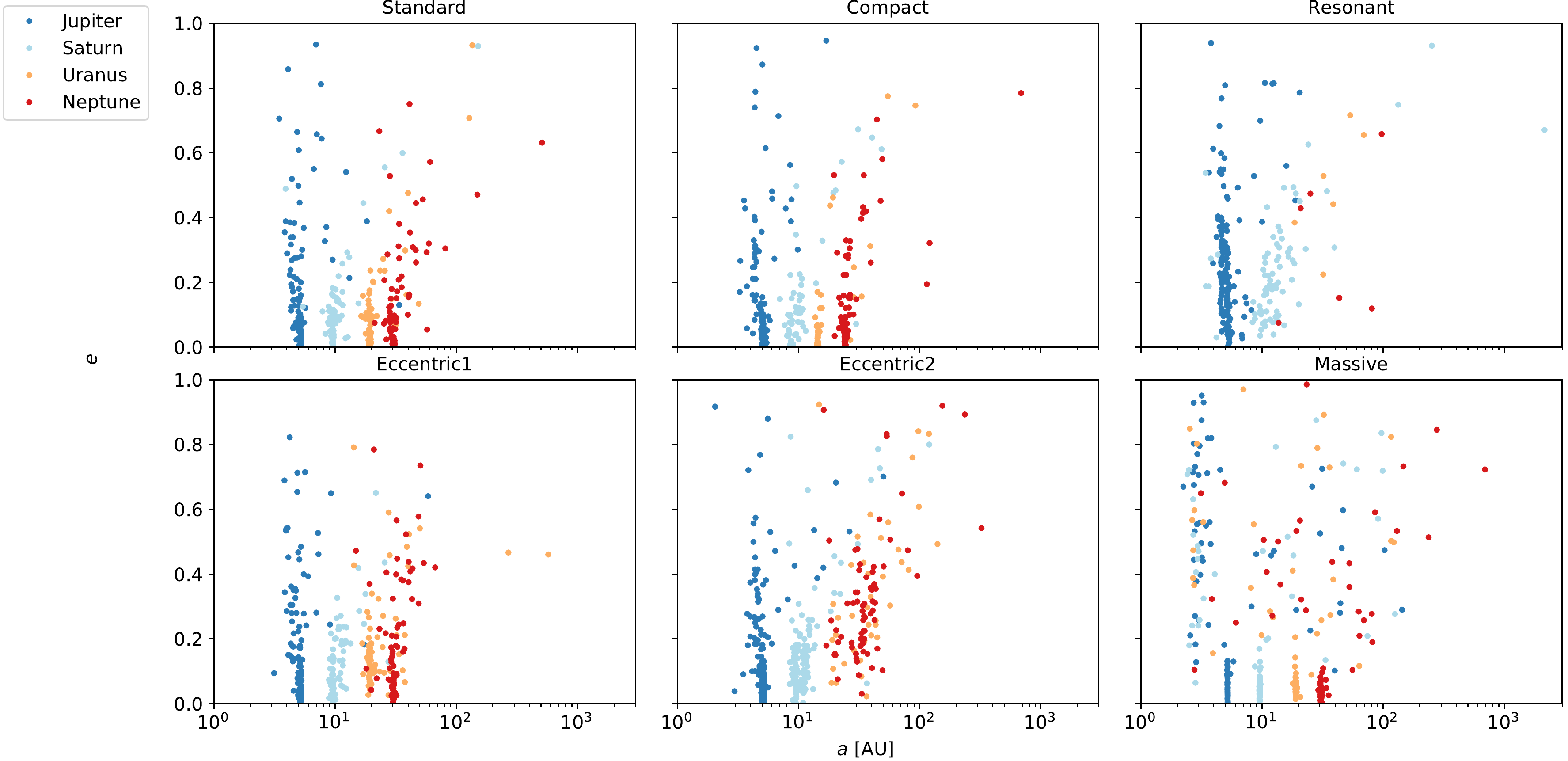}
    \caption{Same as Fig.~\ref{fig:a-e_space_8k} but for the 64k cluster.}
    \label{fig:a-e_space_64k}
\end{figure*}

\begin{figure*}
    \centering
    \includegraphics[width=1\columnwidth]{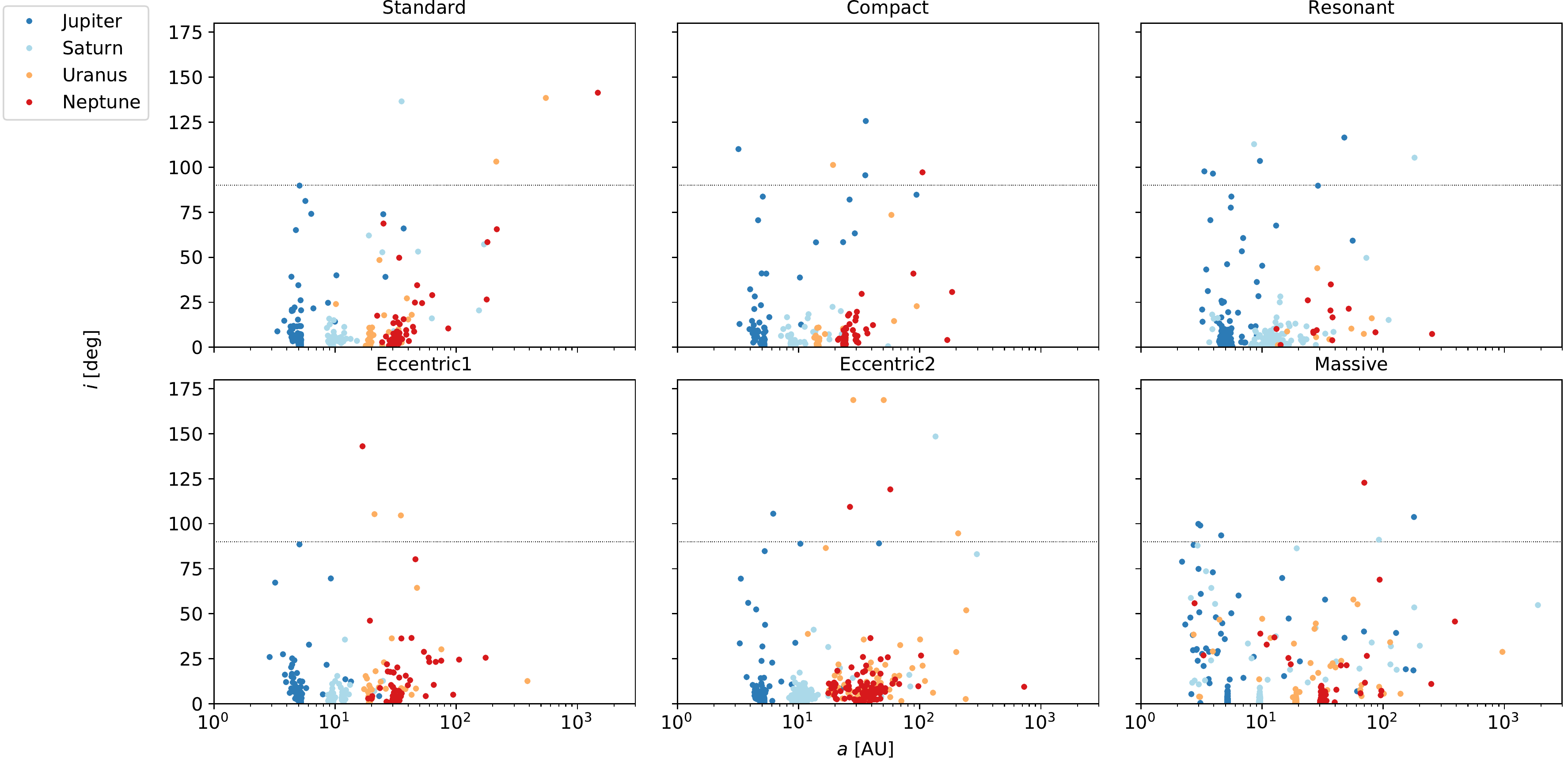}
    \caption{Same as Fig.~\ref{fig:a-i_space_N8k} but for the 16k cluster.}
    \label{fig:a-i_space_N16k}
\end{figure*}

\begin{figure*}
    \centering
    \includegraphics[width=1\columnwidth]{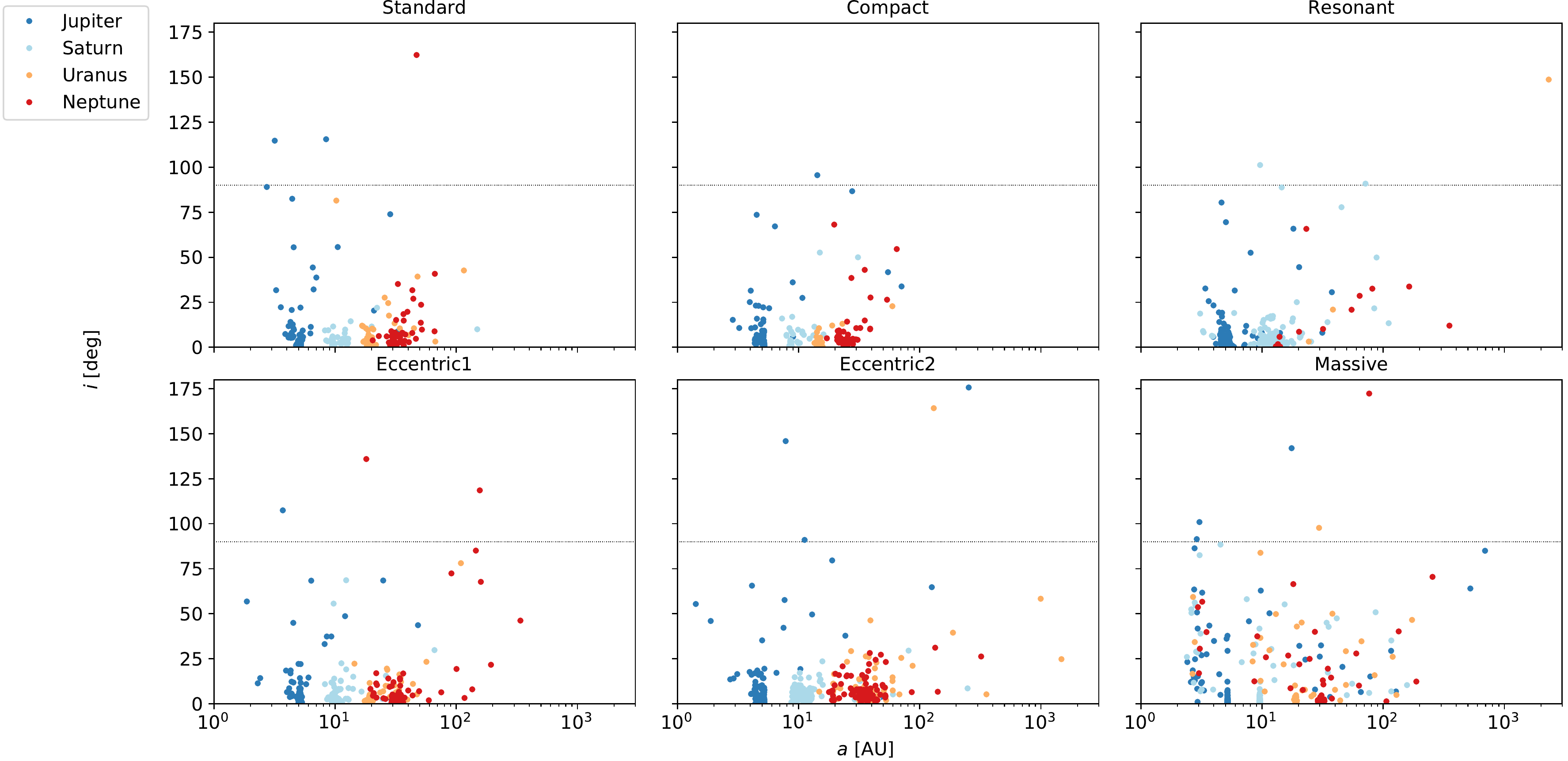}
    \caption{Same as Fig.~\ref{fig:a-e_space_8k} but for the 32k cluster.}
    \label{fig:a-i_space_N32k}
\end{figure*}

\begin{figure*}
    \centering
    \includegraphics[width=1\columnwidth]{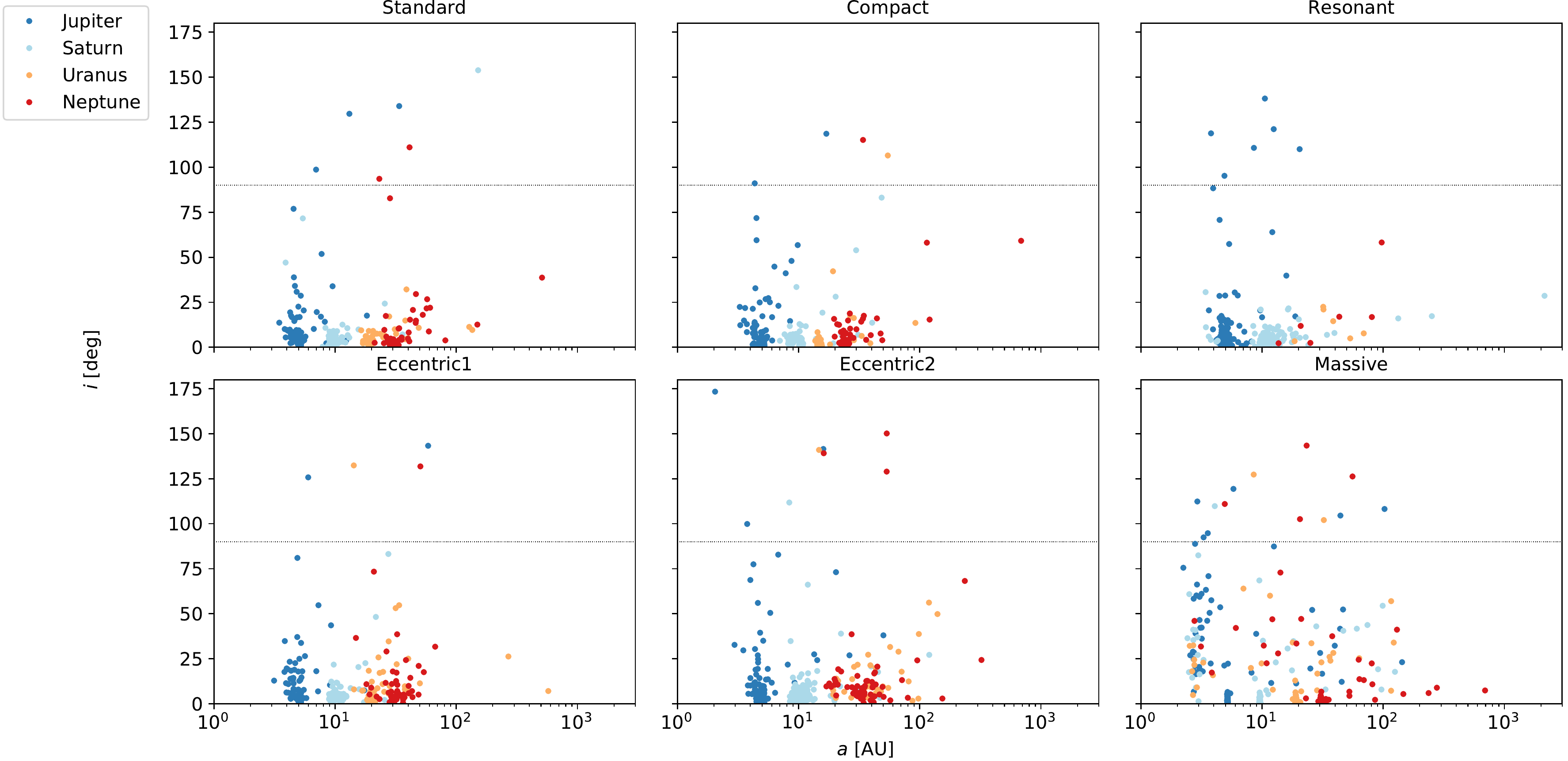}
    \caption{Same as Fig.~\ref{fig:a-e_space_8k} but for the 64k cluster.}
    \label{fig:a-i_space_N64k}
\end{figure*}

\begin{figure*}
    \centering
    \includegraphics[width=1\columnwidth]{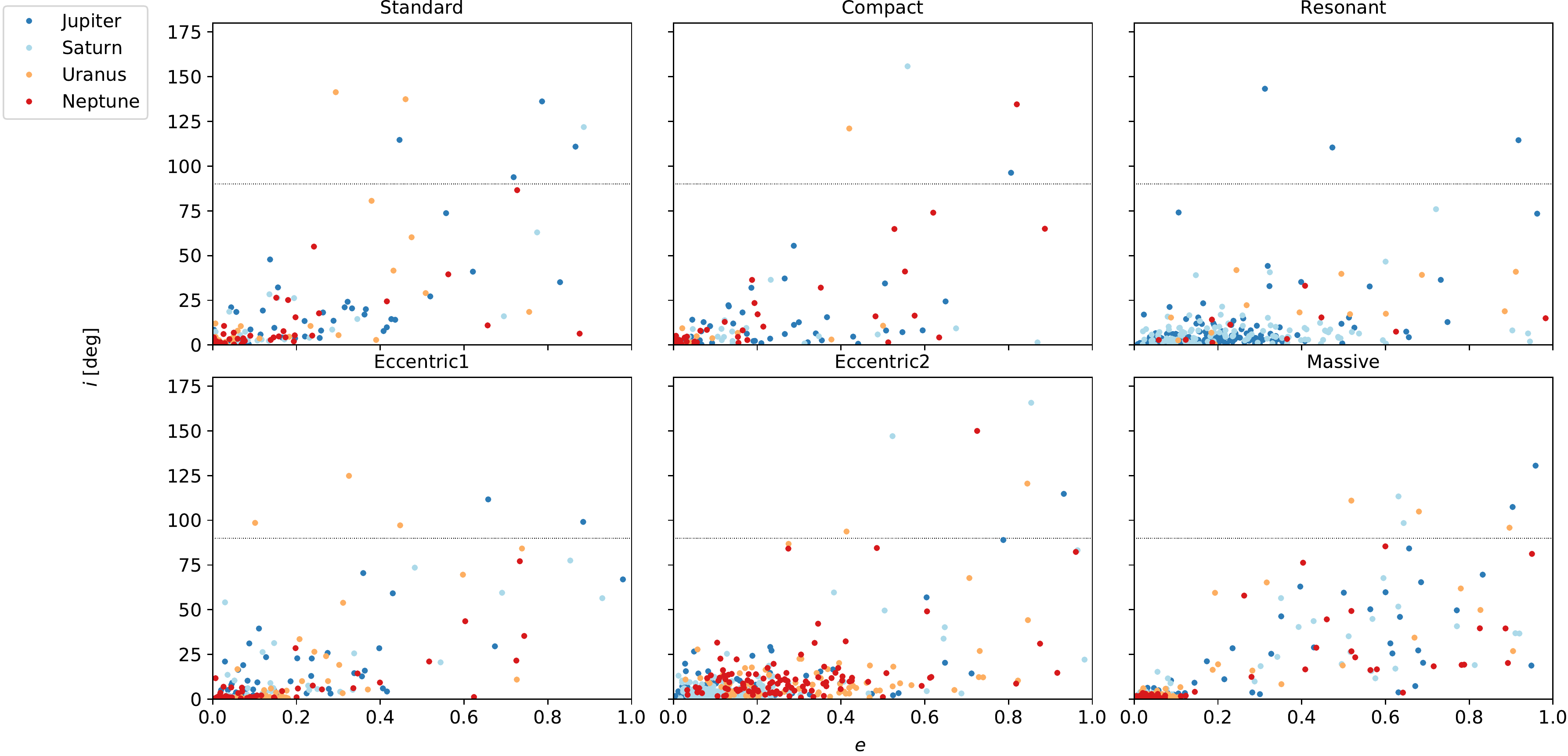}
    \caption{The $e$-$i$ space for all planets in the 8k star cluster which are not ejected from their host planetary system after a simulation time of 100\,Myr for the six different initial configurations. The dotted black shows the threshold of $i=90^\circ$. Planets near that value have polar orbits while those above it have retrograde orbits.}
    \label{fig:e-i_space_N8k}
\end{figure*}

\begin{figure*}
    \centering
    \includegraphics[width=1\columnwidth]{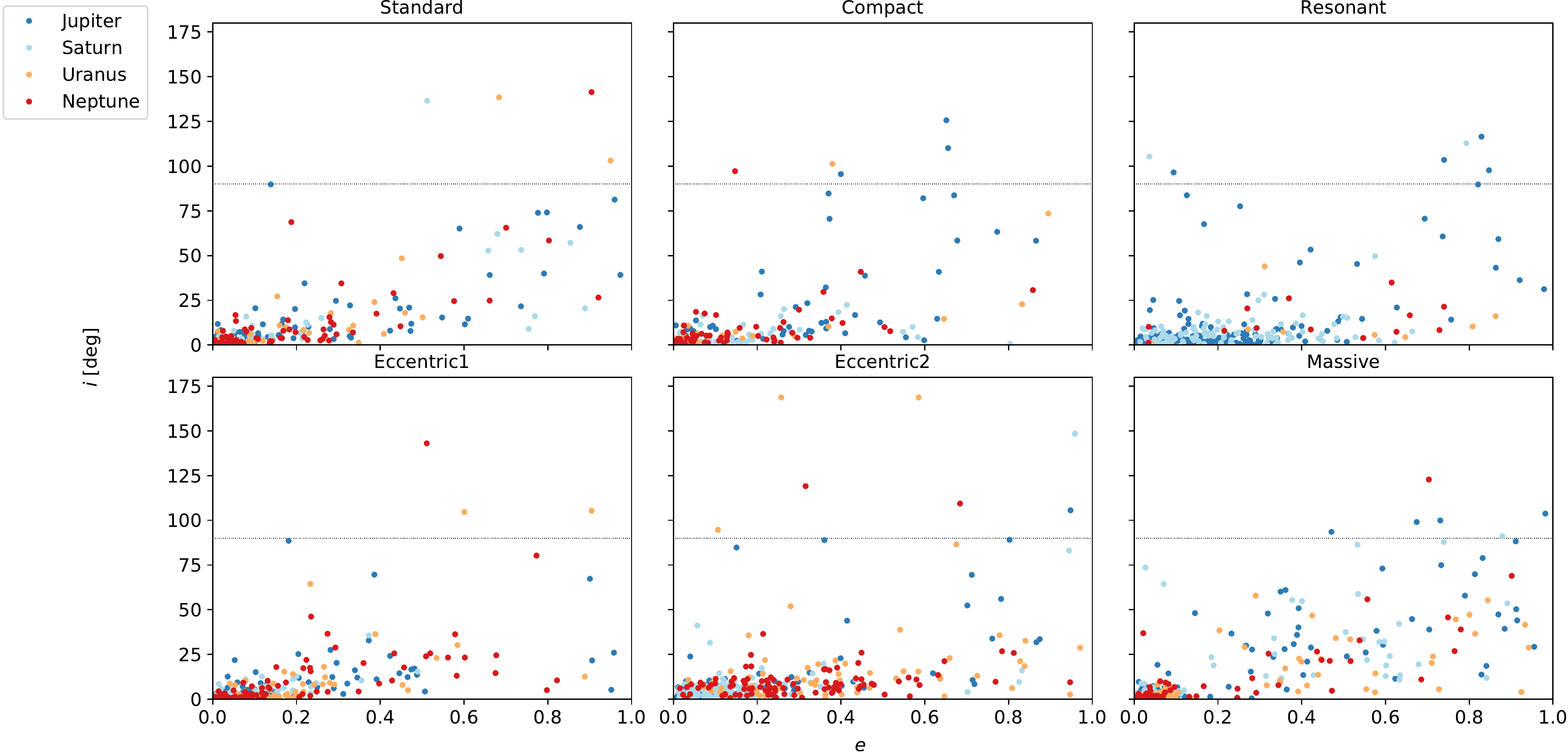}
    \caption{Same as in Fig.~\ref{fig:e-i_space_N8k} but for the 16k cluster.}
    \label{fig:e-i_space_N16k}
\end{figure*}

\begin{figure*}
    \centering
    \includegraphics[width=1\columnwidth]{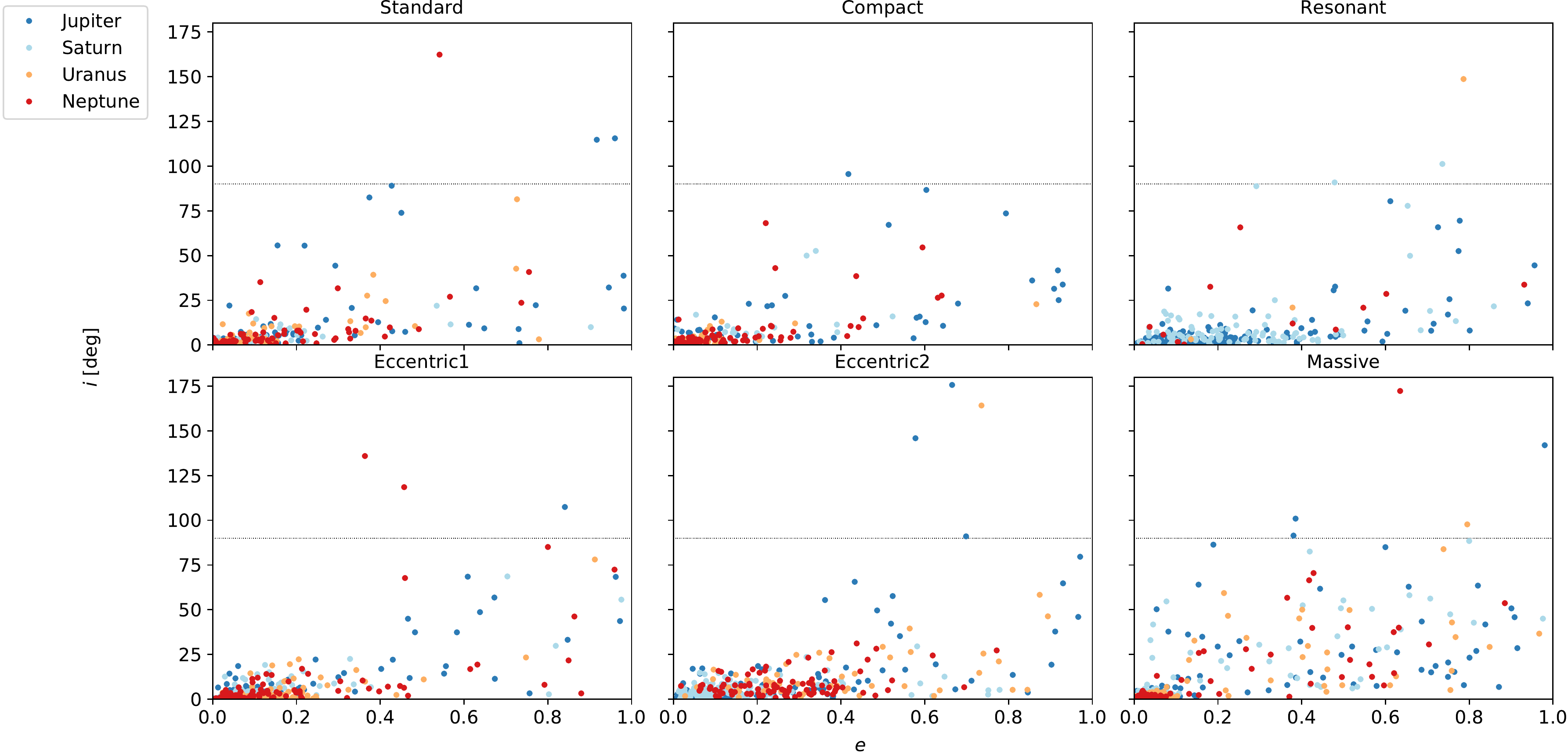}
    \caption{Same as in Fig.~\ref{fig:e-i_space_N8k} but for the 32k cluster.}
    \label{fig:e-i_space_N32k}
\end{figure*}

\begin{figure*}
    \centering
    \includegraphics[width=1\columnwidth]{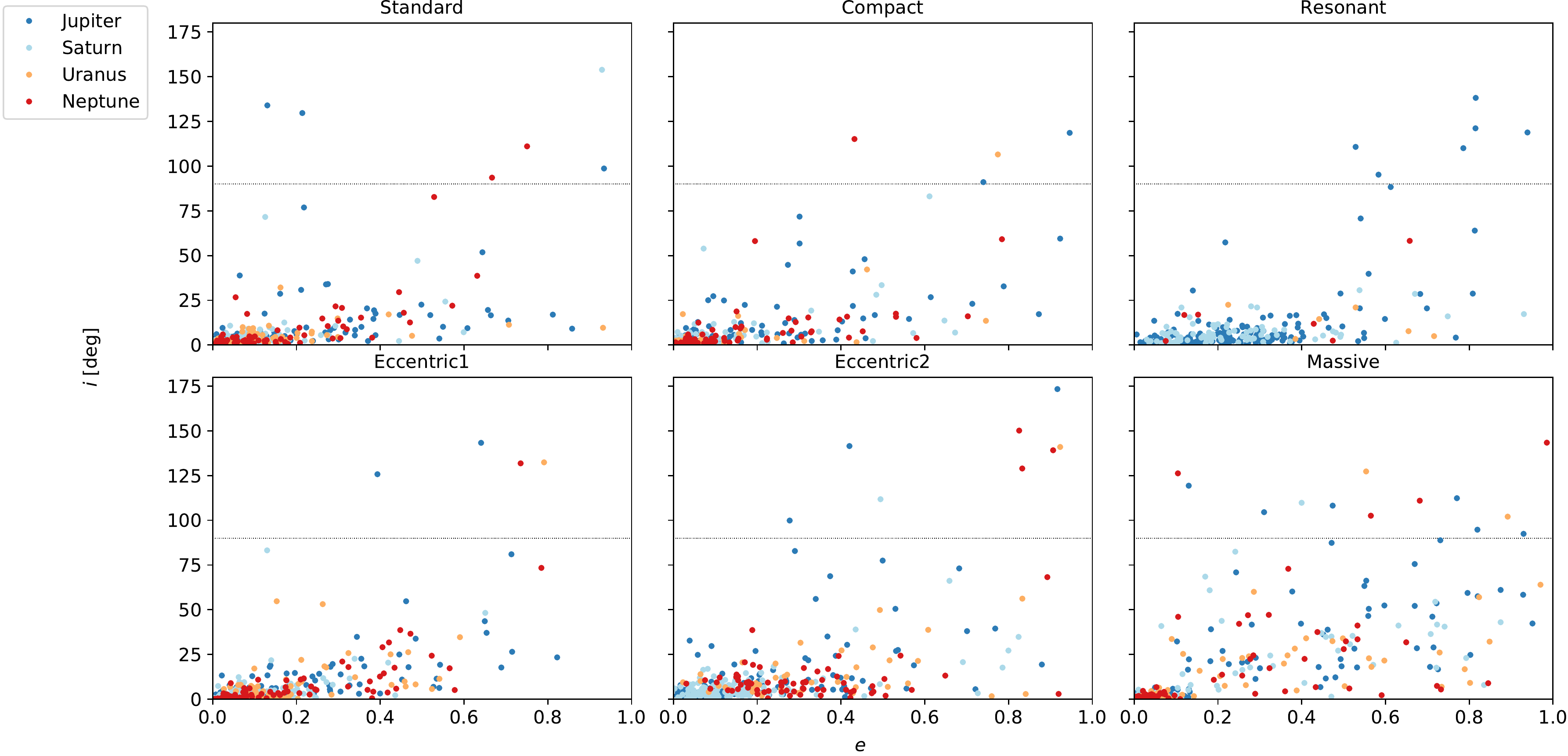}
    \caption{Same as in Fig.~\ref{fig:e-i_space_N8k} but for the 64k cluster.}
    \label{fig:e-i_space_N64k}
\end{figure*}

\bsp
\label{lastpage}
\end{document}